\documentclass[preprint,floatfix]{revtex4}  

\pdfoutput=1

\usepackage{graphicx}
\usepackage{amssymb}
\usepackage{amsmath}

\newcommand{\cf}{{\it cf.}\ }
\newcommand{\eg}{, {\it e.g.},\ }
\newcommand{\ie}{, {\it i.e.},\ }
\newcommand{\etal}{{\it et al. }}
\newcommand{\beq}{\begin{equation}}
\newcommand{\eeq}{\end{equation}}
\newcommand{\bea}{\begin{eqnarray}}
\newcommand{\eea}{\end{eqnarray}}

\newcommand\tab[1]{Table~\ref{tab:#1}}
\newcommand\eqn[1]{(\ref{eq:#1})}

\newcommand\fig[1]{Fig.~\ref{fig:#1}}

\newcommand\sect[1]{Section~\ref{sec:#1}}
\newcommand\figurewidth{7.5cm}

\usepackage{color} % For manuscript iteration

\begin{document}

%%%%%%%%%%%%%%%%%%%%%%%%%%%%%%%%%%%%%%%%%%%%%%%%%%%%%%%%%%%%%%%%%%%%%%%%%%%%%%

\title{Incremental dimension reduction of tensors with random index}

\author{Fredrik~Sandin}     
\email{fredrik.sandin@gmail.com}
\affiliation{EISLAB, Lule{\aa} University of Technology, 971 87 Lule{\aa}, Sweden}  

\author{Blerim~Emruli} 
\affiliation{EISLAB, Lule{\aa} University of Technology, 971 87 Lule{\aa}, Sweden}  

\author{Magnus~Sahlgren} 
\affiliation{Gavagai AB, Sk{\aa}negatan 97, 116 35 Stockholm, Sweden}  

\begin{abstract}
We present an incremental, scalable and efficient dimension reduction technique for tensors that is based on sparse random linear coding.
Data is stored in a compactified representation with fixed size, which makes memory requirements low and predictable.
Component encoding and decoding are performed on-line without computationally expensive re-analysis of the data set.
The range of tensor indices can be extended dynamically without modifying the component representation.
This idea originates from a mathematical model of semantic memory and a method known as random indexing in natural language processing.
We generalize the random-indexing algorithm to tensors and present signal-to-noise-ratio simulations for representations of vectors and matrices.
We present also a mathematical analysis of the approximate orthogonality of high-dimensional ternary vectors,
which is a property that underpins this and other similar random-coding approaches to dimension reduction.
To further demonstrate the properties of random indexing we present results of a synonym identification task.
The method presented here has some similarities with random projection and Tucker decomposition, but it performs well at high dimensionality only ($n>10^3$).
Random indexing is useful for a range of complex practical problems\eg in natural language processing, data mining, pattern recognition, event detection, graph searching and search engines.
Prototype software is provided.
It supports encoding and decoding of tensors of order~$\geq 1$ in a unified framework\ie vectors, matrices and higher order
tensors.
\end{abstract}

\maketitle

%%%%%%%%%%%%%%%%%%%%%%%%%%%%%%%%%%%%%%%%%%%%%%%%%%%%%%%%%%%%%%%%%%%%%%%%%%%%%%

\section{Introduction}
\label{sec:introduction}

The choice of coordinate system often is of great importance when analyzing real-world phenomena and data.
In some cases appropriate basis vectors or basis functions can be chosen from prior knowledge about the system.
If a model does not exist, statistical approaches such as principal component analysis (PCA) or singular value decomposition (SVD) are used.
With an appropriate choice of basis, high-dimensional data often can be represented in a subspace that captures the essential features of the data.
The data analysis process\eg classification, clustering or denoising, is then more efficient and reliable.
The standard statistical approaches for dimension reduction are optimal in the sense that specific statistical features of the data are preserved.
Typical drawbacks of this are the high computational complexity, and that all data needs to be re-analyzed when new data is added or that new data is biased by old estimates.
Nonlinear dimension reduction techniques\eg autoencoder neural networks \cite{Hinton28072006}, do in some cases perform better than the linear methods in terms of class separability, but this comes at the cost of increased computational complexity.

In this paper we present a dimension reduction technique for tensors that enables incremental encoding and decoding of tensor components.
It does not require that the whole data set is stored and no computationally expensive re-analysis is needed when new data is inserted.
Instead, this method encodes the components in a compactified data structure so that decoding of significant features is possible with high probability.
The range of tensor indices can be extended dynamically, without modifying the component representation (the rank is fixed at construction).
This work was stimulated by a method used in natural language processing that is known as {\it random indexing} \cite{kanerva_2000}, see \cite{sahlgren_2005,kanerva_hyperdimensional_2009} for an introduction.
Random indexing is recently applied in a variety of applications, such as indexing of literature databases \cite{vidya_2010}, event detection in blogs \cite{Jurgens_2009} and graph searching for the semantic web \cite{DamljanovicPC2010}.
The idea of random indexing \cite{kanerva_2000} originates from Pentti Kanervas work on sparse distributed memory \cite{SDM_1988}, and related work on the mathematics of brain-inspired information processing with hyperdimensional symbols, see \cite{kanerva_hyperdimensional_2009} for a recent review.
In terms of basis functions this approach is based on the philosophy that ``randomness is the path of least assumption''\ie when little is known beforehand the optimal representation strategy is a random one.
Random indexing of vectors (rank-one tensors) is implemented in the S-Space Package for semantic spaces \cite{Jurgens_2010} and the Semantic Vectors Package \cite{Widdows_2008}.
Here we generalize these approaches to matrices and higher-order tensors in a unified framework:
we outline the equations and properties of {N-way random indexing} (NRI).
The major strengths of this method are that it enables on-line incremental dimension reduction and analysis of complex multi-dimensional data, and that new classes of data can be incorporated without modifying the representation.
The cost of this flexibility is a reduced signal to noise ratio due to the approximate (lossy) nature of the representation scheme, see \sect{simulation}.
Efficient methods for approximate tensor representation are generic and useful tools in the processing of the large amounts of data produced in the modern society.
Higher-order (rank $> 2$) representations are increasingly useful as the processing capacity of computers increases, and allows for more in-depth analyses of complex phenomena.
See \sect{language} for examples.
We restrict this work to spaces with an Euclidean metric, $\delta_{ij}$, so the tensors considered here are practically multi-dimensional arrays.

%%%%%%%%%%%%%%%%%%%%%%%%%%%%%%%%%%%%%%%%%%%%%%%%%%%%%%%%%%%%%%%%%%%%%%%%%%%%%%

\subsection{Related work}
\label{sec:relatedwork}

This work was stimulated by Pentti Kanervas work on sparse distributed memory \cite{SDM_1988} and a number of papers related to hyperdimensional computing and neuro-symbolic computing, see \cite{kanerva_hyperdimensional_2009} for a recent review.
In particular it has bearing on, and is a generalization of the random indexing method developed for natural language processing \cite{kanerva_2000,sahlgren_2005}, see also p.~153 in \cite{kanerva_hyperdimensional_2009} where Kanerva mentions the possibility to extend random indexing to two dimensions.
Random projection \cite{Papadimitriou_1998} and random mapping \cite{Kaski_1998} are other methods that utilize high-dimensional random vectors for dimension reduction.
Random projection is used to reduce the dimensionality of a set of points in Euclidean space, while approximately preserving pairwise distances.
This possibility follows from the Johnson--Lindenstrauss lemma \cite{johnson_1984}, which essentially states that a small set of points in high-dimensional space can be mapped into a space of lower dimension such that the distances between the points are approximately preserved.
Random projection has enabled several breakthrough developments\eg in the field of algorithms and in elegant alternative proofs.
It is used for combinatorial optimization, machine learning and problems related to information retrieval, see \cite{VempalaBook_2004} for a review.
The method described here is based on random high-dimensional vectors and projection operators also, but the mathematical structure of the algorithm is different from that of the random projection method, and the properties of the technique from an application point of view are different.
Our method enables dimension reduction of tensors of any rank (rank $\leq 3$ is feasible today on a PC).

There are a few well-known algorithms for dimension reduction of tensors, see \cite{TensorReview} for a recent review.
In particular the Tucker decomposition and the parallel factor model are commonly used methods, and there are some extensions of these two algorithms that are commonly used also \cite{TensorReview}.
Tucker decomposition (known also as N-mode PCA, N-mode SVD etc.) is a form of higher-order principal component analysis,
which decomposes a tensor, $c_{ijk\ldots}$, according to the scheme
\beq
	c_{ijk\ldots} = \sum_{\alpha}\sum_{\beta}\sum_{\gamma}\ldots~g_{\alpha\beta\gamma\ldots}~a_{i\alpha}b_{j\beta} c_{k\gamma}\ldots,
	\label{eq:tucker}
\eeq
where $g_{\alpha\beta\gamma\ldots}$ is the core tensor and $\{a_{i\alpha},~b_{j\beta},~c_{k\gamma}\ldots\}$ are factor matrices.
The factor matrices are usually taken to be orthogonal and can be thought of as the principal components in each dimension of the tensor.
The core tensor represents interactions between the principal components.
Several methods to compute the Tucker decomposition have been developed, see \cite{TensorReview} for a review.
These methods are computationally expensive, because the goal to find an ``optimal'' decomposition with side constraints on the factor matrices is non trivial.
The method presented in this paper is somewhat similar to the Tucker decomposition, because it uses a mathematically equivalent expression for decoding of the representation.
Our method is essentially different from Tucker decomposition in the sense that it is based entirely on random coding, and it thereby avoids the computationally expensive process of calculating the factor matrices.
In our model the factor matrices are randomly generated objects, so-called random indices.
Another benefit is that our method is incremental, but it performs well only with large tensors (high dimensionality is a prerequisite).
In contrast, Tucker decomposition is useful for low-dimensional problems due to the computational complexity of the method.
The parallel factor decomposition is a special case of the Tucker decomposition, which results if the core tensor in \eqn{tucker} is enforced to be superdiagonal.

%
% http://en.wikipedia.org/wiki/Compressed_sensing
%

%%%%%%%%%%%%%%%%%%%%%%%%%%%%%%%%%%%%%%%%%%%%%%%%%%%%%%%%%%%%%%%%%%%%%%%%%%%%%%

\section{Indifference property of high-dimensional spaces}
\label{sec:highd}

High-dimensional spaces are different from the two- and three-dimensional spaces that we are naturally trained to imagine.
In two and three dimensions, randomly generated vectors of equal norm are rather similar\eg when compared with the dot product.
This is not so at high dimensionality, where nearly all vectors are unsimilar.
Another counter-intuitive property is that the volume of the unit hypersphere relative to that of the hypercube with corresponding width rapidly approaches zero at increasing dimensionality.
The tendency of vectors in high-dimensional binary spaces to be different is well described in \cite{SDM_1988}.
We introduce some key points from that work here and then we will show how to generalize these ideas to ternary space $\{-1,0,1\}^n$, which is important in this work.
Consider the binary space $\{0,1\}^n$ of vectors with length $n$ and equal probability of the states $0$ and $1$.
The distance, $d$, between two binary vectors can be defined as the number of non-zero bits in the bit-wise exclusive or ({\it xor}) of the vectors.
This is equivalent to the square of the Euclidean distance and it corresponds to how many bits that are different in two vectors.
The number of vectors in the space that are at a distance $d$ from a specific vector is given by the binomial coefficient
\beq
	C(n,d) = \binom{n}{d},
\eeq
because this is the number of different ways to choose (flip) $d$ bits out of $n$.
The number of vectors at a certain distance from a reference vector therefore follows the binomial distribution with probability $p=1/2$, which has mean $n/2$ and variance $n/4$.

At high values of $n$ the binomial distribution can be approximated with a normal distribution.
If a distribution is approximately normal the proportion within $z$ standard deviations of the mean is erf$(z/\sqrt{2}).$
This implies that the distance distribution is highly concentrated around the mean, because the error function quickly approaches unity for increasing $z$.
For example, $99.7$\% of the distances are within three standard deviations from the mean distance.
Only one billionth $(10^{-9})$ of the distances deviate more than six standard deviations from the mean.
The mean distance is $n/2$ and the standard deviation of distance is $\sqrt{n}/2$.
This implies that the mean distance is $\sqrt{n}$ standard deviations\eg $31.6$ standard deviations for $n=1000$.
A striking consequence of this distribution of distances is that practically all vectors in a high-dimensional binary space are located at a distance $\sim n/2$ from any specific vector.
For $1000$-bit vectors the standard deviation is $\sqrt{1000}/2\simeq 15.8$~bits.
Six standard deviations correspond to $95$~bits.
All but one billionth of $1000$-bit vectors are located at $500 \pm 95$~bits from {\it any} specific vector in that space.
The concentration of distances around the mean increases with $n$ and it implies that randomly generated high-dimensional vectors are {\it indifferent} with high probability.
The indifference property is at the core of the ideas discussed in Pentti Kanervas book on sparse distributed memory \cite{SDM_1988}, and related work on the mathematics of brain-inspired information processing with hyperdimensional symbols, see \cite{kanerva_hyperdimensional_2009} for an introduction.
Concepts and symbols represented with arbitrary random high-dimensional patterns are indifferent by chance, but new associations and transformations can be constructed from existing representations with a learning mechanism.
Next we consider indifference (orthogonality) in high-dimensional ternary space.

%%%%%%%%%%%%%%%%%%%%%%%%%%%%%%%%%%%%%%%%%%%%%%%%%%%%%%%%%%%%%%%%%%%%%%%%%%%%%%

\section{Properties of high-dimensional ternary space $\{-1,0,1\}^n$}
\label{sec:ternary}

% "I often reflect that had the Ternary instead of the binary Notation been adopted
% in the Infancy of Society, machines something like the present would long ere this
% have been common, as the transition from mental to mechanical calculation would
% have been so very obvious and simple". (Thomas Fowler, 1840)

In this work we are interested in ternary vectors\ie instead of bits with two possible states, $\{0,1\}$, we consider (balanced) trits with states $\{-1,0,1\}$.
The introduction of a state with negative sign is important, we will return to that in the following section.
This discussion concerns ternary vectors of length $n$ with $k$ positive ($1$) and $k$ negative ($-1$) states, where $k\ll n/2$\ie we are interested in sparse ternary vectors with vanishing sum.
The ternary space can be visualized as a subset of an inner product space where orthogonality is defined by a vanishing dot product between two vectors.
With this (the usual) definition of orthogonality, it follows that an $n$-dimensional ternary space has at most $n$ mutually orthogonal vectors.
However, in a high-dimensional space there are many more vectors that are ``nearly orthogonal''.
This is analogous to the high probability of indifference between vectors in high-dimensional binary space, which is described above.
In the following we justify this essential point with explicit results for the approximate orthogonality of sparse vectors in high-dimensional ternary space.
As far as we know the result has not been presented elsewhere.
  
The total number, $N$, of ternary vectors of length $n$ that has $k$ positive and $k$ negative components is
\beq
	N = \binom{n}{2k}\binom{2k}{k} = \binom{n}{k}\binom{n-k}{k},
	\label{eq:nternary}
\eeq
because there are $C(n,2k)$ different ways to choose $2k$ non-zero states and
$C(2k,k)$ different ways to distribute the signs to the non-zero states.
The alternative (second) definition above can be interpreted in a similar way.
There are $C(n,k)$ different ways to choose the positive states and $C(n-k,k)$ ways to choose the negative states, or vice versa.
These two definitions are mathematically equivalent.
How many of these $N$ vectors have a dot product that is nearly zero\ie how many of them are approximately orthogonal?
Let $d=|\langle\cdot,\cdot\rangle|$ be the absolute value of the dot product between two vectors.
For simplicity we restrict the analysis to $0\leq d\leq k$, because we are interested in approximately orthogonal vectors only.
This restriction does not affect the accuracy of the result.
We assume also that the vectors are sparse so that $n \gg k$.
Imagine a fixed reference vector that is picked at random from the space of $N$ vectors.
This reference vector has $k$ positive states, $k$ negative states and $n-2k$ states that are zero.
The large majority of vectors with $\langle\cdot,\cdot\rangle=\pm d$ with respect to this reference vector will have $d$ states that coincides with the $2k$ non-zero states of the reference vector, and the remaining $2k-d$ non-zero states will be distributed among the $n-2k$ states that are zero in the reference vector.
There are additional vectors with the same value of $d$, because cancellations of type $1+1-1=1$ result from higher-order coincidences.
The relative number of such vectors is, however, insignificant and we therefore neglect them here. This simplification is justified with a numerical calculation, we will return to that below.
The selection of $2k-d$ non-zero states out of $n-2k$ gives a factor of $C(n-2k,2k-d)$.
Then remains the question how many possibilities there are to select those $2k-d$ non-zero states from the $2k$ non-zero states in the reference vector, and how many combinations that arise because of signs.
These questions are not independent, because the number of ways to choose $2k-d$ states from $2k$ states depends on the number of $+1$ states that are chosen, and the relative number of $+1$ states that are chosen will affect also the number of possible permutations.
Accounting for these constraints the number of vectors is
\begin{widetext}
\beq
	N(n,k,d) \simeq
	\binom{n-2k}{2k-d}
	\sum_{n_+=k-d}^{k} \binom{k}{n_+}\binom{k}{2k-d-n_+}\binom{2k-d}{n_+},
	\quad d\leq k,~n\gg k,
	\label{eq:ninner}
\eeq
\end{widetext}
where $n_+$ denotes the number of positive states that are chosen from the $2k$ non-zero states in the reference vector.
The number of negative states chosen is $n_- = 2k-d-n_+$.
The sum in \eqn{ninner} arises because there are multiple choices for the number of positive states to choose from the reference vector.
At most $k$ positive states can be chosen\ie all positive states.
The lower limit of $n_+=k-d$ corresponds to the maximum value for the number of negative states chosen, $n_-=k$.
The first factor in the sum, $C(k,n_+)$, accounts for the number of ways to choose $n_+$ positive states from the $k$ positive states in the reference vector.
Similarly, the second factor accounts for the number of ways to choose $n_-$ negative states from the $k$ negative states in the reference vector.
The last factor accounts for sign permutations when distributing the chosen states to the $2k-d$ non-zero states that are selected by the prefactor.
Since the number of positive and negative signs are fixed, the combinatorial problem solved here should have a hypergeometric character.
This is indeed the case, because the sum in \eqn{ninner} can be replaced with a generalized hypergeometric function,
\begin{widetext}
\bea
	N(n,k,d) &\simeq&
	\binom{n-2k}{2k-d}
	\binom{2k-d}{k}
	\binom{k}{k-d} \nonumber \\
	&& \times ~~ {}_3F_2(-d,-k,-k;~1+k-d,1+k-d;~-1),
	\quad d\leq k,~n\gg k.
	\label{eq:ninnerhyp}
\eea
\end{widetext}
The generalized hypergeometric function, $_3F_2$, is a standard mathematical function that is described\eg on-line and in the book \cite{NIST_book}.

If we divide the number of vectors, $N(n,k,d)$, which has a specific value of $d$ with respect to any reference vector, with the total number of vectors in the space, $N$, the result is the relative size of the space as a function of $d$.
The relative size of the space is equivalent to the probability of randomly choosing a vector from the space that has a dot product of $\pm d$ with respect to a reference vector,
\beq
	P(n,k;~\langle\cdot,\cdot\rangle=\pm d) \simeq N^{-1} N(n,k,d), \quad d\leq k,~n\gg k.
	\label{eq:pinner0}
\eeq
This distribution function is the result that we are looking for, because it describes the probability that randomly chosen vectors from the space are nearly orthogonal.
The numbers $N$ and $N(n,k,d)$ are enormous ($n$ is a high number).
For practical purposes we therefore make a series expansion of factors involving $n$ in the limit $n\rightarrow\infty$.
The result is,
\begin{widetext}
\bea
	%%%%%%%%%%%%%%%%%%%%%%%%%%%%%%%%%%%%%%%%%%%%%%%%%%%%%%%%%%%%
	% Taylor expansion of factors containing n
	P(n,k;~\langle\cdot,\cdot\rangle=\pm d) &\simeq&
	\frac{T_1 + T_2}{n^d}
	\sum_{i=0}^{d}\frac{(k!)^4}{
	\left[(k-d+i)!\right]^2
	\left[(k-i)!\right]^2
	(d-i)!~i!} \nonumber \\
	&=& 
	\frac{(T_1 + T_2)d!}{n^d}
	\binom{k}{d}^2
	{}_3F_2(-d,-k,-k;~1+k-d,1+k-d;~-1),
	\label{eq:pinner1} \\
	%%%%%%%%%%%
	T_1 &=& 1-\frac{8k^2+d^2+d-8kd}{2n}, \\
	T_2 &=& \frac{1}{n^2}\left[
	2(1-2k)^2 k^2
	+ \frac{d^4}{8}
	+ \left(\frac{5}{12}-2k\right)d^3
	+ \left(10k^2-4k+\frac{3}{8}\right)d^2 \right. \nonumber \\
	&& \left. +\left(-16k^3+10k^2-2k+\frac{1}{12}\right)d
	\right] + {\cal O}(n^{-3}),
\eea
\end{widetext}
where the terms $T_1$ and $T_2$ originate from the series expansion.
The assumptions $d\leq k$ and $n\gg k$ are to be respected in applications of this result.
Numerical results for the dot product between a reference vector and $10^{12}$ randomly chosen ternary vectors are presented in \tab{dotp}.
\begin{table}[h]
\caption{Approximate orthogonality of the high-dimensional space $\{-1,0,1\}^n$.
Tabulated here is the probability, $P$, in \eqn{pinner1} for different values of the vector length, $n$, and number of non-zero components, $2k$.
These probabilities are to be compared with the corresponding probabilities obtained from explicit numerical simulations, $P_{sim}$.
Entries marked with an asterisk demonstrate the effect of neglecting contributions to the inner product arising from higher-order trit combinations (like $\langle\cdot,\cdot\rangle = \ldots+1\times 1\ldots-1\times 1\ldots+1\times 1\ldots=1$) in the analysis leading to \eqn{pinner1}.
The series expansion is marginally applicable in the case $n=10^2$ for low values of $k$, and $n \gg k$ is violated for high $k$.
}
\vspace{1ex}
\centering
\begin{tabular}{| r | r | l  r | l r | l r |}
\hline
& & \multicolumn{2}{|c|}{$n=10^2$} & \multicolumn{2}{|c|}{$n=10^3$} & \multicolumn{2}{|c|}{$n=10^4$} \\[0.5ex]
$2k$ & $\langle\cdot,\cdot\rangle$ & 
$~P_{sim}$ & $P~~~$ & 
$~P_{sim}$ & $P~~~$ & 
$~P_{sim}$ & $P~~~$ \\[0.5ex]
\hline
4  &      0 & 8.5e-1 & \quad  8.47e-1& 9.8e-1	& \quad 9.84e-1& $\sim$1.0 & \quad 9.98e-1 \\
   & $\pm$1 & 7.3e-2 & 7.29e-2& 7.9e-3 & 7.93e-3& 8.0e-4 & 7.99e-4\\
   & $\pm$2 & 2.0e-3 & 1.94e-3& 2.0e-5 & 1.99e-5& 2.0e-7 & 2.00e-7\\
\hline
8  &      0 & 5.5e-1 & *5.17e-1& 9.4e-1 & 9.38e-1& 9.9e-1 & 9.94e-1\\
   & $\pm$1 & 1.9e-1 & 1.90e-1& 3.0e-2 & 3.05e-2& 3.2e-3 & 3.20e-3\\
   & $\pm$2 & 2.8e-2 & 2.74e-2& 3.9e-4 & 3.86e-4& 4.0e-6 & 3.99e-6\\
   & $\pm$3 & 2.0e-3 & 1.96e-3& 2.4e-6 & 2.44e-6& 2.5e-9 & 2.49e-9\\
   & $\pm$4 & 7.4e-5 & 7.42e-5& 8.2e-9 & 8.22e-9& $<$$10^{-10}$ & 8.3e-13\\
\hline
12 &      0 & 3.5e-1 & --~~~~& 8.7e-1 & 8.65e-1& 9.9e-1 & 9.86e-1\\
   & $\pm$1 & 2.3e-1 & --~~~~& 6.4e-2 & 6.37e-2& 7.1e-3 & 7.10e-3\\
   & $\pm$2 & 7.9e-2 & --~~~~& 2.0e-3 & 1.99e-3& 2.2e-5 & 2.17e-5\\
   & $\pm$3 & 1.6e-2 & --~~~~& 3.4e-5 & 3.44e-5& 3.7e-8 & 3.69e-8\\
   & $\pm$4 & 2.0e-3 & --~~~~& 3.6e-7 & 3.64e-7& $<$$10^{-10}$ & 3.8e-11\\
\hline
16 &      0 & 2.5e-1   & --~~~~& 7.8e-1 & *7.73e-1& 9.7e-1 & 9.75e-1\\
   & $\pm$1 & 2.0e-1   & --~~~~& 1.0e-1 & 1.02e-1& 1.3e-2 & 1.25e-2\\
   & $\pm$2 & 1.1e-1   & --~~~~& 5.9e-3 & 5.94e-3& 7.1e-5 & 7.09e-5\\
   & $\pm$3 & 4.3e-2 & --~~~~& 2.0e-4 & 2.01e-4& 2.3e-7 & 2.34e-7\\
   & $\pm$4 & 1.1e-2 & --~~~~& 4.4e-6 & 4.44e-6& 4.9e-10 & 5.0e-10\\
\hline
20 &      0 & 2.0e-1 & --~~~~& 6.9e-1 & *6.72e-1& 9.6e-1 & 9.61e-1\\
   & $\pm$1 & 1.8e-1 & --~~~~& 1.4e-1 & 1.39e-1& 1.9e-2 & 1.93e-2\\
   & $\pm$2 & 1.2e-1 & --~~~~& 1.3e-2 & 1.31e-2& 1.8e-4 & 1.75e-4\\
   & $\pm$3 & 6.5e-2 & --~~~~& 7.4e-4 & 7.36e-4& 9.6e-7 & 9.55e-7\\
   & $\pm$4 & 2.7e-2 & --~~~~& 2.8e-5 & 2.78e-5& 3.5e-9 & 3.49e-9\\
\hline
\end{tabular}
\label{tab:dotp}
\end{table} 
These numerical results confirm the analytical result.
Observe, however, that the accuracy of the analytical result is poor for low values of $n$ and high values of $k$, as indicated in the table.
This is connected to the assumption that $n\gg k$ in the analysis above. Note the prefactor of $n^{-d}$ in the series expansion, which implies that the probability for high values of $d$ is low.

\begin{figure*}[h]
\centering
\begin{minipage}[b]{7.5cm}
\centering
\includegraphics[width=\linewidth]{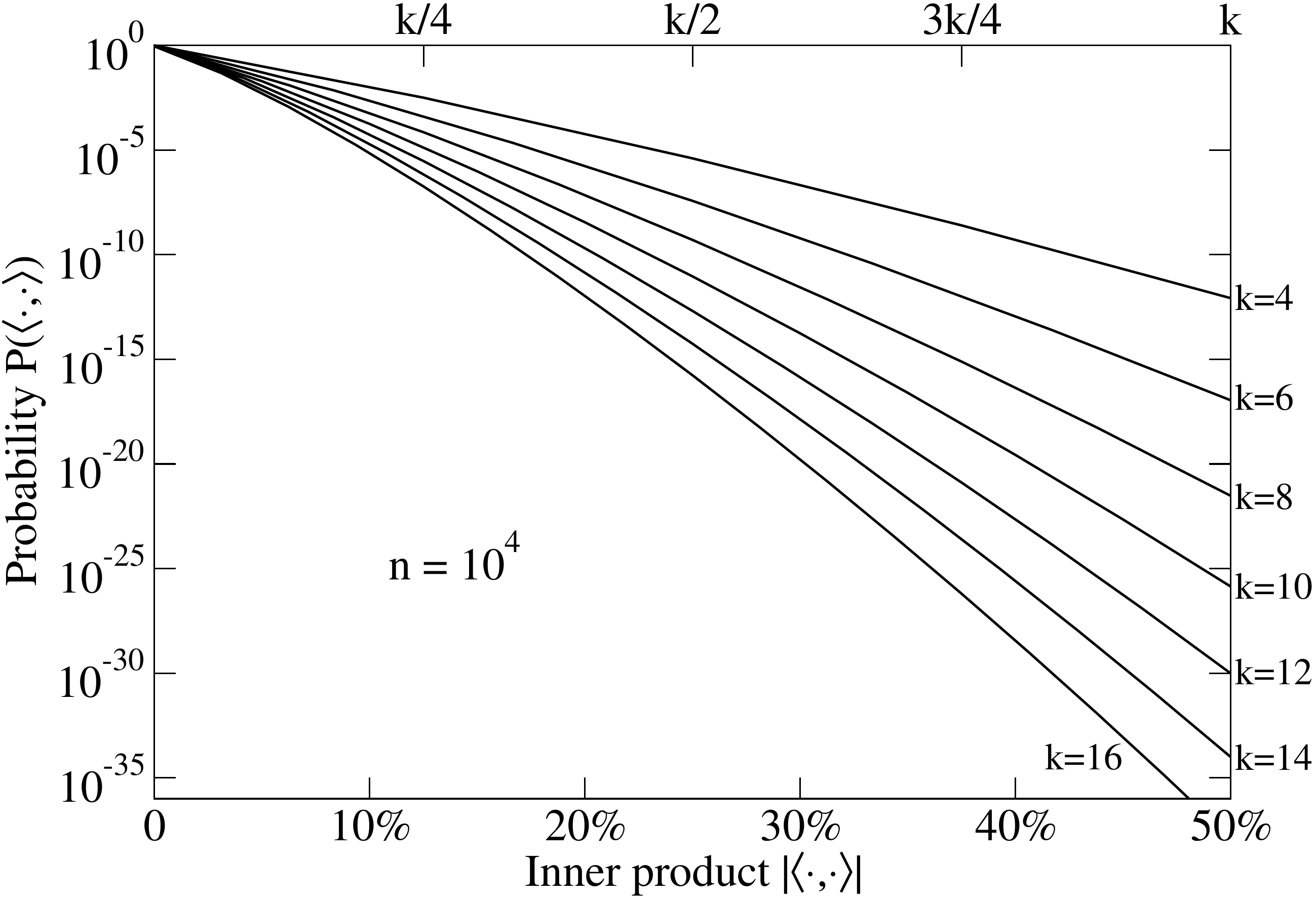}
\end{minipage}
\hspace{0.5cm}
\begin{minipage}[b]{7.5cm}
\centering
\includegraphics[width=\linewidth]{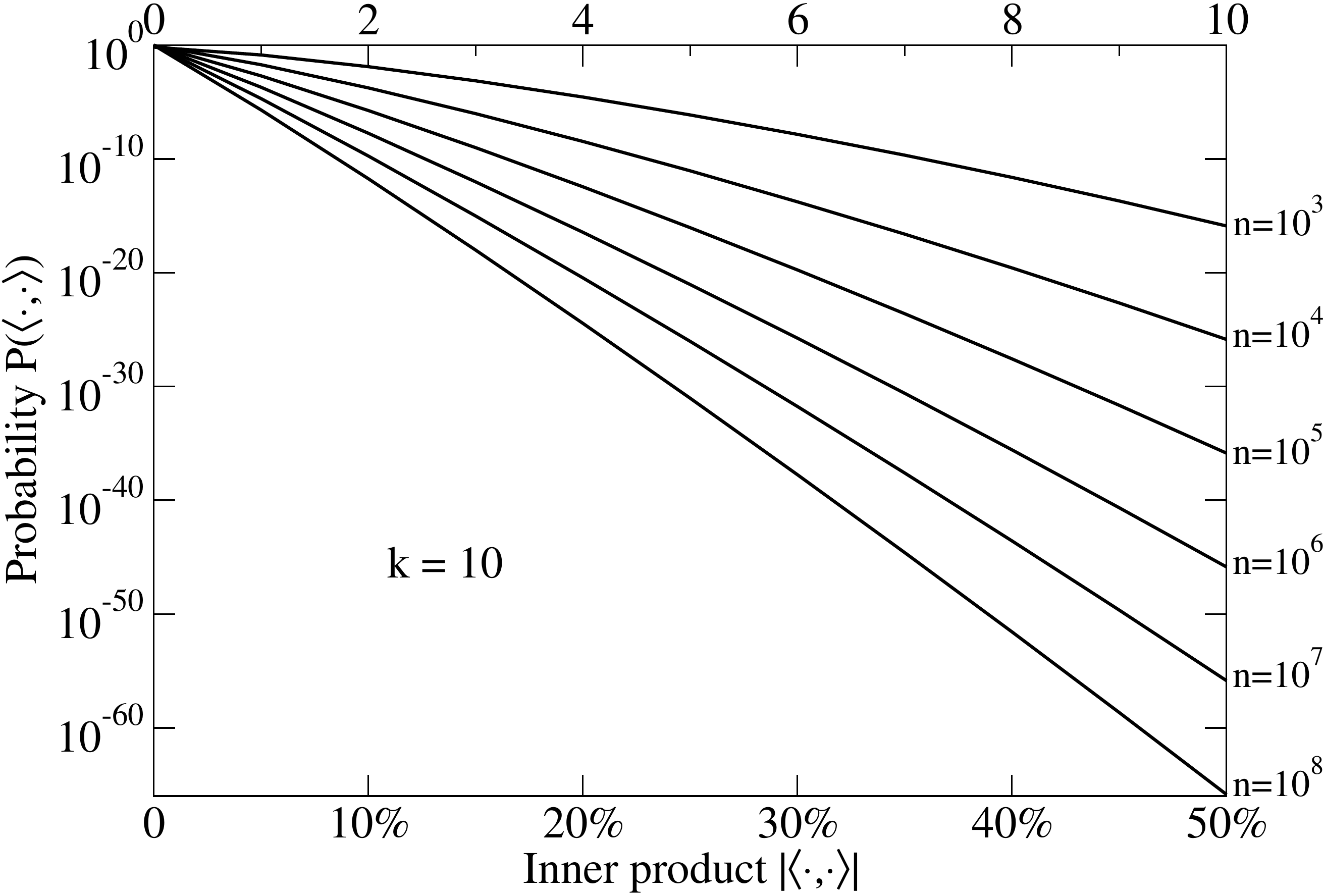}
\end{minipage}
\caption{
Approximate orthogonality of high-dimensional ternary space $\{-1,0,1\}^n$. 
The panel on the left-hand side shows the probability \eqn{pinner1} for inner products of sparse ternary vectors of length $n=10^4$ and different numbers of non-zero components, $2k$.
The panel on the right-hand side shows the probability \eqn{pinner1} for $k=10$ and different lengths of the ternary vectors, $n$.
In both cases the horizontal scale is normalized to the maximum value of the inner product, which is $2k$.
Probabilities for absolute values of $\langle\cdot,\cdot\rangle$ higher than $50$\% of the maximum are excluded, because \eqn{pinner1} is valid for $d\leq k$ only.
For $n=10^4$ and $k=4$\ie ternary vectors of length ten thousand with four positive and four negative trits, the probability that a randomly generated vector has an inner product of four with respect to a reference vector is about $10^{-12}$.
The probability of an inner product of minus four is about $10^{-12}$ also.
Similarly, for $n=10^4$ and $k=12$ the probability of $50$\% overlap ($\langle\cdot,\cdot\rangle = \pm 12$) is about $10^{-30}$.
Note the prefactor $n^{-|\langle\cdot,\cdot\rangle|}$ in the series expansion of the probability \eqn{pinner1}, which indicates that the probability of high absolute values of the inner product is low.
}
\label{fig:dotp}
\end{figure*}

For historical reasons we consider the example $n=10^4$ and $k=10$, which is typical in Kanervas work \cite{SDM_1988}.
It then follows from \tab{dotp} that $96$\% of the space is orthogonal with respect to a reference vector, and less than $4$\% ($2\times 1.93$) of the space has a dot product of $+1$ or $-1$.
Only $7\times 10^{-9}$ of the space has a dot product with a magnitude higher than or equal to four, which corresponds to $\sim 20$\% non-zero trits in common.
With $25$\% common trits ($d=5$ and $k=10$) the relative size of the space is $2\times 10^{-11}$.
This demonstrates that most of the space is approximately orthogonal to any particular vector in the space.
Analogously, the dot product of pairs of vectors that are randomly chosen from the space follows the same distribution\ie equation \eqn{pinner1}.
The probabilities for $n=10^4$ and some different values of $k$ are illustrated in \fig{dotp}. 
Note that for a given subset of $N_s$ vectors the distribution of dot products is different, because there are more pairs of vectors than individual vectors in a set.
This is similar to birthday type problems, where the probability that at least two people in a room has the same birthday depends on the number of pairs rather than the number of individuals in the room.
A similar situation arise in the context of hash tables, where collisions are more frequent than suggested by the number of hash values.
The number of pairs in a subset of $N_s$ vectors is $C(N_s,2)=N_s(N_s-1)/2$.
This implies that the number of pairs increases roughly as the square of the number of vectors in the set, and that the number of vectors that can be chosen randomly with a low probability for significant correlations should be reduced with a square root compared to what is suggested by equation \eqn{pinner1}, \tab{dotp} and \fig{dotp}.
Next we present the generalized random indexing algorithm, which is based on this idea of approximate orthogonality.

%%%%%%%%%%%%%%%%%%%%%%%%%%%%%%%%%%%%%%%%%%%%%%%%%%%%%%%%%%%%%%%%%%%%%%%%%%%%%%

\section{Random-index coding of tensors}
\label{sec:ricoding}

The concept of random indexing (RI) is introduced in \sect{introduction}.
It was invented for dimension reduction and semantic analysis in the context of natural language processing.
In that context the RI method is used to encode vectors that represent the ``meaning'' of words, so-called semantic vectors, or context vectors.
Here we generalize RI to tensors of arbitrary rank and present the algorithm of N-way random indexing (NRI).

In the following, tensor components are denoted with $c_{ijk\ldots}$ and indices $\{i,~j,~k,\ldots\}$ are used in component space.
Tensor {\it states} are denoted with $s_{\alpha\beta\gamma\ldots}$ and indices $\{\alpha,~\beta,~\gamma,\ldots\}$ are used in state space.
The state tensor has a physical representation that is stored in memory, but it is accessed by encoder and decoder functions only.
Tensor components, $c_{ijk\ldots}$, are related to the states and constitute the input to (output from) the encoder (decoder) function.
%In other words, states $s_{\alpha\beta\gamma\ldots}$ are stored in memory and the components $c_{ijk\ldots}$ are I/O channels.
The rank of the state tensor is equivalent to that of $c_{ijk\ldots}$, but the state tensor may be of significantly smaller size.
For each index (dimension) of these tensors there is an associated random-index tensor, $r_{{\cal D},i\alpha}$, where ${\cal D}$ is a dimension index.
For vectors ${\cal D}=1$, for matrices ${\cal D} \in \{1,2\}$ and so on.
%This means that vectors have one associated object of type $r_{1,i\alpha}$, while matrices have two, $r_{1,i\alpha}$ and $r_{2,i\alpha}$, and so on.
For given values of ${\cal D}$ and $i$ the state-space components of $r_{{\cal D},i\alpha}$ form a sparse high-dimensional random-index vector, referred to here as {\it index vectors}: 
\beq
	r_{{\cal D},i\alpha} = [\ldots~0~0~0~1~0~0~0~\ldots~0~0~0~{-1}~0~0~0\ldots]_{{\cal D},i}.
\eeq
Index vectors have a few non-zero components at random positions\ie at random values of $\alpha$ (thereby the name ``random index'').
The non-zero components have an absolute value of one and exactly half of them are negative\ie this is a sparse balanced ternary vector, see \sect{ternary}
(note that the symbol $k$ in that section is different from the index $k$ defined here).
The number of non-zero components, $\chi_{\cal D}$, is a model parameter, which has a typical value of order ten.
We denote the ranges of state indices, $\{\alpha,~\beta,~\gamma,\ldots\}$, with $[1,n_{\cal D}]$ so that\eg $\alpha \in [1,n_1]$ and $\beta \in [1,n_2]$.
Similarly, the ranges of component indices, $\{i,~j,~k,\ldots\}$, are $[1,N_{\cal D}]$.
The length of an index vector is equivalent to the maximum value of the state index, $n_{\cal D}$, in each dimension.
For example, if the state tensor is of size 1000x2000 the index vectors would be of length 1000 (2000) for ${\cal D}=1$ (${\cal D}=2$).
Observe that index vectors can be represented in compact form\eg as illustrated in \fig{encoding}, because most components are zero.
An effective way of doing that is to store the indices of non-zero components only.
Signs can be encoded implicitly with the position of the indices\eg first half are positive.
The number of non-zero components in an index vector is denoted with $\chi_{\cal D}$, which is an even number.
For each dimension, ${\cal D}$, there are $N_{\cal D}$ index vectors of length $n_{\cal D}$ and each index vector has $\chi_{\cal D}$ non-zero components.
In practical applications index vectors are represented in compact form by at most a few dozen integers, so the storage space required for an NRI representation is essentially determined by the size of the state tensor.
%Index vectors of a particular dimension have equal $l^2$ norm and the sum of components is zero.
A summary of parameters and definitions is presented in \tab{summary}.
\begin{table*}
\caption{Summary of parameters and definitions.
}
\vspace{1ex}
\centering
\begin{tabular}{l | l}
\hline
Expression  & Description \\[0.5ex]
\hline
$c_{ijk\ldots}$ & Tensor components \\
$s_{\alpha\beta\gamma\ldots}$ & State tensor, accessed by encoder/decoder functions only \\
${\cal D}$ & Dimension index ($1 \leq {\cal D} \leq$ rank) \\
$N_{\cal D}$ & Number of index vectors in dimension ${\cal D}$, $\{i,~j,~k,\ldots\} \in [1,N_{\cal D}]$ \\
$n_{\cal D}$ & Length of index vectors in dimension ${\cal D}$, $\{\alpha,~\beta,~\gamma,\ldots\} \in [1,n_{\cal D}]$ \\
$\chi_{\cal D}$ & Number of non-zero components in index vectors of dimension ${\cal D}$ \\
${\cal N}=\prod\chi_{\cal D}$ & Number of states that encode one tensor component \\
$\propto\prod n_{\cal D}$ & Disk/memory space required to store the state tensor \\
$\propto\sum N_{\cal D}\chi_{\cal D}$ & Disk/memory space required to store index vectors \\
\hline
\end{tabular}
\label{tab:summary}
\end{table*}

\subsection{Encoding algorithm}

\begin{figure*}[h]
\centering
\includegraphics[width=9cm]{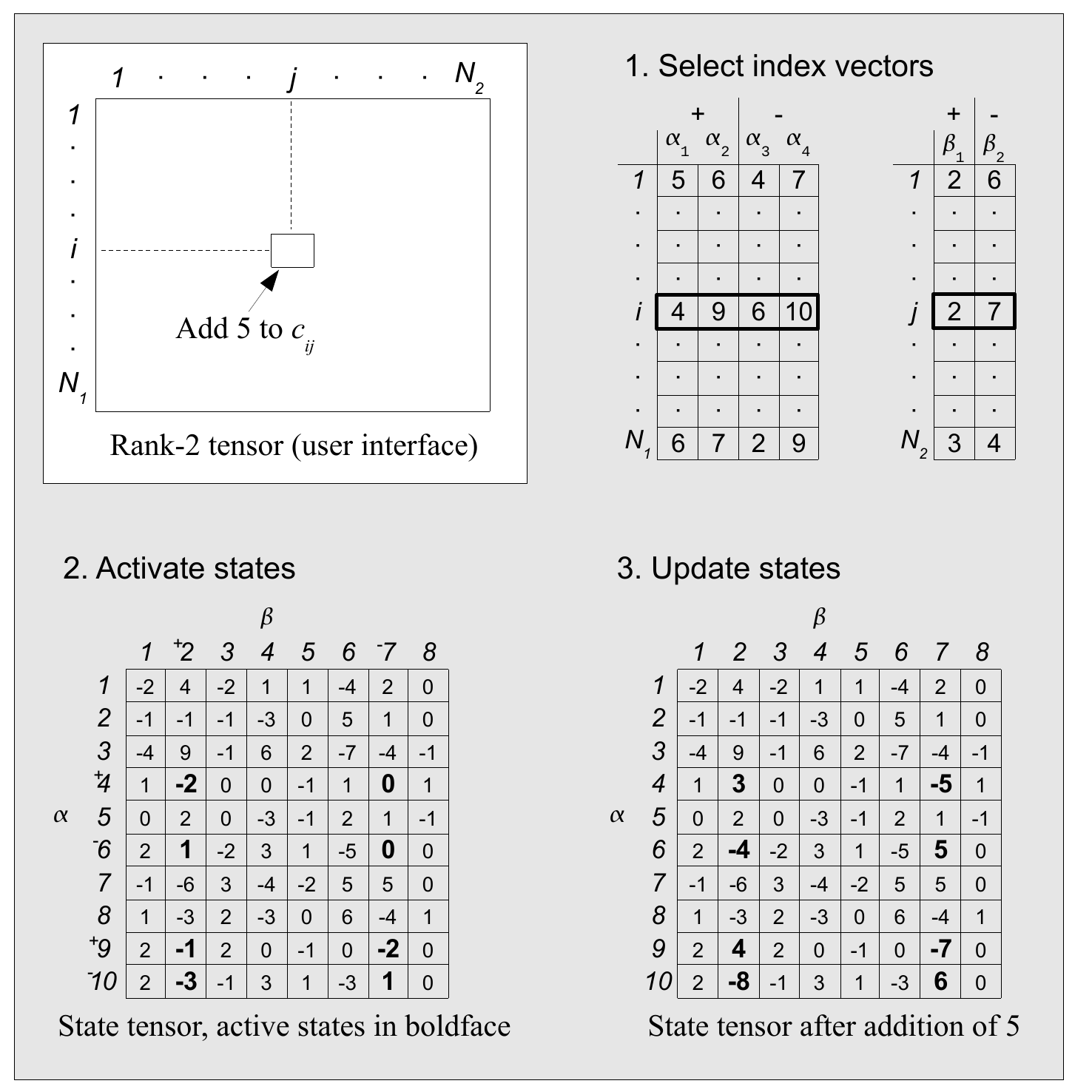}
\caption{Encoding of a tensor component with two-way random indexing.
The highlighted rank-two tensor exists in user space\ie users can write and read it's component values.
It is a virtual object without physical representation, which is connected to the physical datastructures illustrated in the shaded area via encoder and decoder functions.
In this example a value of $5$ is added to the tensor component $c_{ij}$ of a rank-two tensor of size $N_1 \times N_2$.
The three-step encoding procedure is illustrated in the shaded area.
First the indices $i$ and $j$ of the component are used to select the appropriate index vectors.
The second step is to activate the subset of states that encode the value of $c_{ij}$.
Note that (the sign of) the activated rows and columns correspond to (the sign of) the indices in the index vectors.
The third and last step is to add $5$ to the selected states, while respecting the sign combination of the activations.
A few remarks are in order.
This method performs well for high-dimensional state tensors only\ie for high values of all $n_{\cal D}$ (the rank can be low or high).
Index vectors correspond to sparse ternary vectors $\{-1,0,1\}^n$ with vanishing sum, which are represented in compact form by storing the indices of non-zero components only and grouping the indices with respect to their sign.
The three-step process illustrated here is mathematically equivalent to \eqn{encoding}.
}
\label{fig:encoding}
\end{figure*}

State components, $s_{\alpha\beta\gamma\ldots}$, are initially set to zero.
This implies that the tensor components $c_{ijk\ldots}$ are zero also (see decoding).
After initialization, the components are updated with addition and subtraction operations, not by assignment.
This is necessary because assignment has unwanted side effects on other components due to the random nature of the representation.
In other words, multiple components are affected when one component is modified and assignment destroys statistics.
A tensor component $c_{ijk\ldots}$ is encoded in the state tensor $s_{\alpha\beta\gamma\ldots}$ using the index vectors.
The addition of a scalar weight $w$ to a tensor component $c_{ijk\ldots}$ corresponds to the operation
\beq	
	s_{\alpha\beta\gamma\ldots} \rightarrow s_{\alpha\beta\gamma\ldots}
	 + w(r_{1,i\alpha}~r_{2,j\beta}~r_{3,k\gamma}~\ldots),
	\label{eq:encoding}
\eeq
where the indices $\{i,~j,~k,\ldots\}$ are fixed by the choice of tensor component.
This means that the indices of the tensor component are used to select a particular set of index vectors, forming a subset of approximately orthogonal vectors in state space.
The outer product of index vectors is a tensor with a few ($\cal N$) non-zero components with values $+1$ and $-1$.
It has the same rank and size as the state tensor.
Subtraction of $w$ is defined by the replacement $w \rightarrow -w$ in \eqn{encoding}.
The encoding process is illustrated in \fig{encoding} using compact representation of index vectors.

\subsection{Decoding algorithm}

The decoding operation is a projection of the state tensor on the index vectors corresponding to the tensor component $c_{ijk\ldots}$
\beq
	c_{ijk\ldots} = {\cal N}^{-1}\sum_{\alpha,\beta,\gamma\ldots}
	r_{1,i\alpha}~r_{2,j\beta}~r_{3,k\gamma}~\ldots~s_{\alpha\beta\gamma\ldots},
	\label{eq:decoding}
\eeq
where $\cal N$ is a normalization factor that is to be defined below.
The encoding procedure \eqn{encoding} is based on a sequence of outer products of index vectors, and the decoding procedure is the corresponding sequence of inner products, which results in a projection of the state tensor on the index vectors.
An essential aspect of this dimension reduction technique is the approximate orthogonality of the high-dimensional index vectors, see \sect{ternary}.
The approximate orthogonality makes it possible to decode many of the significant components of $c_{ijk\ldots}$ from the compressed and distributed representation in the state tensor.
For index vectors of length $n_{\cal D}$, at most $n_{\cal D}$ linearly independent vectors can be constructed (a set of basis vectors).
However, equation \eqn{pinner1} and \fig{dotp} illustrates that for large $n_{\cal D}$ there are many more vectors that are approximately orthogonal.
This makes it possible to encode and decode tensor components in a distributed and approximately orthogonal subspace of the state tensor, which is only partially overlapping the subspaces representing other tensor components.
In practice the result is that tensor components $c_{ijk\ldots}$ with high accumulated weight often can be identified, while tensor components with low accumulated weight disappear in noise.
A practical implementation of the decoding function may therefore return a top-list of components $c_{ijk\ldots}$ with high values, for given values of some of the indices $\{i,~j,~k,\ldots\}$.
These high-value components are to be interpreted as ``likely significant features''.
We return to the details of this approach in the next section, which deals with simulation results.
Note the similarity between the decoding operation \eqn{decoding} and the Tucker decomposition \eqn{tucker}.

The normalization factor in \eqn{decoding} compensates for the sum over states\ie the redundancy caused by adding $w$ to multiple states in the encoding procedure.
An index vector in dimension $\cal D$ has $\chi_{\cal D}$ non-zero components.
The normalization factor therefore is
\beq
	{\cal N} = \prod_{\cal D} \chi_{\cal D}.
	\label{eq:normalization}
\eeq
For example, the rank-two tensor in \fig{encoding} has ${\cal N} = 4\times 2 = 8$, because the index vectors have, respectively, four and two non-zero components.
The quantity $\cal N$ is a measure of the computational complexity of the encoding and decoding processes, because $\cal N$ states are accessed when encoding or decoding the value of one single tensor component.
\begin{figure*}[h]
\centering
\includegraphics[width=9cm]{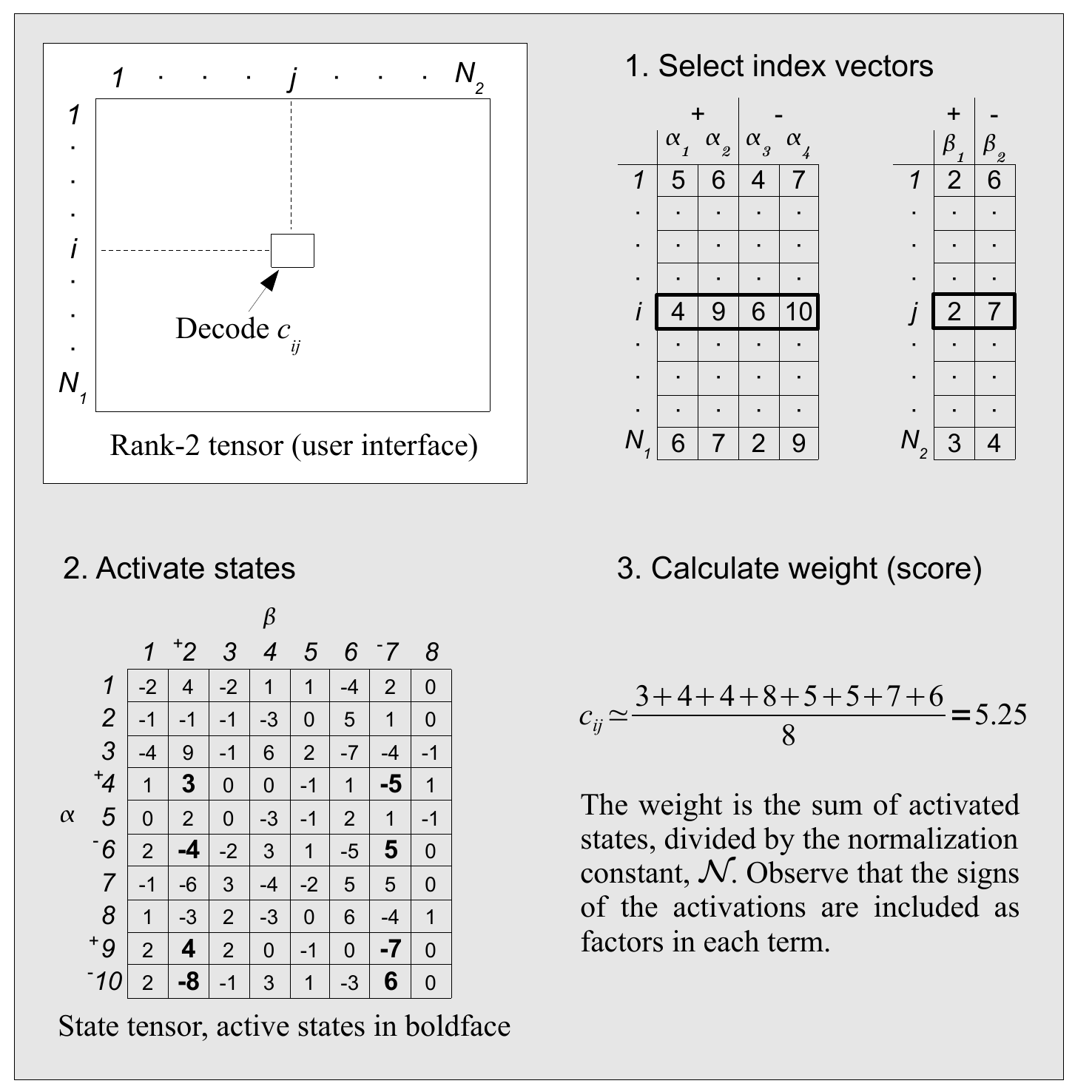}
\caption{Decoding of a rank-two tensor component with the random-index method.
The process illustrated here is the inverse of that in \fig{encoding}.
Refer to that figure for further details.
The first step is to select the appropriate index vectors using the indices $i$ and $j$ of the tensor component that is to be decoded.
The index vectors are then used to activate the subset of states that encode the value of $c_{ij}$.
Note that (the sign of) the activated rows and columns correspond to (the sign of) the indices in the index vectors.
The third and last step is to sum the activated states, taking the overall sign into account, and divide the sum with the normalization constant \eqn{normalization}.
Note that the decoded value is similar to that encoded in \fig{encoding}.
The difference ($5.25$ vs $5.00$) is a consequence of the non-zero initial states.
This three-step process is mathematically equivalent to \eqn{decoding}.
}
\label{fig:decoding}
\end{figure*}

\subsection{One-way random indexing}
\label{sec:oneway}

The RI approach used in natural language processing is based on rank-one tensors \cite{kanerva_2000,sahlgren_2005}\ie vectors.
We will return to that in \sect{simulation}, but a  brief description of this special case is included here for completeness.
In the traditional RI algorithm, each word type that appears in a text corpus is associated with a context vector, and each context ({\it e.g.}, document) is associated with a ternary index vector.
A context vector corresponds to the states of a rank-one tensor, and the index vectors are the ternary index vectors of that tensor.
It would be impractical to construct one state tensor with associated index vectors for each class or item that is to be represented.
Therefore, rank-one NRI tensors can be implemented as a set of rank-one tensors sharing the same index vectors.
For rank~$\geq 1$ the relative size of the index vectors, $(\Sigma N_{\cal D} \chi_{\cal D})/(\Pi n_{\cal D})$, is low and a special solution of that type is not necessary.

\subsection{Summarizing remarks}

The N-way random indexing (NRI) method that is outlined above is a linear dimension reduction technique that is to be used with large-size tensors ($n_{\cal D}>10^3$), because the performance is poor at low dimensionality of the index vectors.
This constraint is related to the probability that randomly generated ternary vectors are approximately orthogonal \eqn{pinner1}, see also \fig{dotp}. 
NRI enables computationally efficient incremental dimension reduction of large tensors.
Quantitative details are lost in this process, but qualitative features can be maintained.
This key feature is demonstrated with simulations in the next section, and it follows from a property of the ternary index vectors.
We mentioned in \sect{ternary} that the co-existence of positive and negative states is essential\ie that index vectors should be ternary and not binary.
The reason is that, on average, each state is equally likely to be incremented and decremented when random or non-systematic data is encoded.
Random noise does therefore not saturate the states.
It is only via systematic updates of a subset of tensor components that the states gain relatively high amplitudes.
It is only these high-value components that are likely to be decoded correctly with a high weight, while most low-value components will have low weights that are below the noise threshold.
In this perspective NRI-encoded tensor components can be imagined as the momentaneous values of temporal integrals, and only those integrals that have accumulated relatively high values are likely to be correctly decoded and identified.
The simulation results presented in \sect{simulation} illustrates this point.

The incremental feature of this dimension reduction technique is obvious.
A tensor component can be encoded or decoded at any time by updating or reading $\cal N$ states.
No computationally expensive re-calculation of the basis is needed after the addition of data.
Another incremental feature is that the ranges of tensor indices, $N_{\cal D}$, can be increased dynamically by adding more randomly generated index vectors.
This has a low impact on the memory footprint, because index vectors are small entities and the relatively large state tensor is not modified in this process.
No computationally expensive analysis is needed to construct the index vectors, because these vectors are generated randomly.
The order of tensor component updates does not matter, which practically follows from the commutative property of the addition and subtraction operators.
To break this symmetry\eg for the purpose of coding time evolution or order, one has to introduce an additional tensor index for that degree of freedom.
Temporal random indexing \cite{Jurgens_2009} is an example, which is used to analyze the time evolution of word semantics to detect novel events in on-line texts.

Note that the decoding operation \eqn{decoding} and the distributed representation of tensor components in the state tensor is mathematically similar to the Tucker decomposition \eqn{tucker}.
Tucker decomposition is a generalization of PCA to tensors of rank~$\geq 3$ (the original work addressed rank-three tensors only) and that approach is therefore computationally expensive, and typically limited to smaller tensors than those addressed by NRI.
The state tensor defined here is analogous to the ``counters'' of Kanervas distributed memory model, which symbolizes synapse weights of neurons receiving signals from address-decoder neurons, see Chapter 4 in \cite{SDM_1988}.
The sparse character of the coding scheme used here is common in biological nervous systems, where sparse codes are essential for enegy-efficient information processing.
These analogues are mentioned here for inspiration only, they are not meant to be interpreted literally.
The distributive character of the encoding operation \eqn{encoding} and the projective character of the decoding operation \eqn{decoding} resembles the structure of neuro-symbolic encoding and decoding operations, such as those in Kanervas ``spatter code'' and Tony Plates ``holographic reduced representation'', see \cite{kanerva_hyperdimensional_2009} for references and an overview.
Next, we present simulation results that demonstrate some key properties of one- and two-way RI.

%%%%%%%%%%%%%%%%%%%%%%%%%%%%%%%%%%%%%%%%%%%%%%%%%%%%%%%%%%%%%%%%%%%%%%%%%%%%%%

\section{Numerical study of properties}
\label{sec:simulation}

The randomly generated index vectors used to access the state tensor via outer (distributive) and inner (projective) products are ``nearly orthogonal'' by chance, but not strictly orthogonal, see \fig{dotp}.
Different tensor components are therefore encoded in partially overlapping subspaces of the state tensor, and do therefore interfere with each-other to some degree.
The magnitude of this effect depends on the tensor rank and order of NRI, the length of the index vectors, $n_{\cal D}$, the number of non-zero trits in the index vectors, $\chi_{\cal D}$, the dimension reduction, $\Pi_{\cal D} N_{\cal D}$~:~$\Pi_{\cal D} n_{\cal D}$, and the characteristics of the data that is to be encoded.
In general, tensors with few significant components of comparable magnitude are well represented, while non-sparse tensors with many significant components are poorly represented with NRI.
Components with high values have a negative impact on the reliability of components with significantly lower values.
It is therefore important to control the relative magnitude of significant components.
Observe that the tensor does not have to be sparse to get good performance.
A full tensor can be well represented if it has relatively few high-value components that represent significant features, while the majority of components have low value and are insignificant.
This property is useful if high-dimensional data is accumulated over time and the significant components are unknown beforehand.
In this section we illustrate these aspects with simulations and present an application to natural language processing that further demonstrates the NRI approach.
All simulations are performed with the RITensor software \cite{RITensor}, which includes a C++ template that implements the N-way RI algorithm and some utility functions\eg file I/O.
The template has user-defined data types for the state tensor and index vectors, and a mechanism to monitor eventual state saturation.
RITensor includes a Matlab interface also, which is convenient for small-scale experiments.

We consider an example where a dense rank-two tensor\ie a matrix, is represented with one-way and two-way RI.
Recall that a rank-two tensor can be represented with one-way RI if each column (or row) is treated as a rank-one tensor\ie a vector.
Each column of the matrix represents a class and each row represents a feature that the classes may have.
In principle the matrix elements (tensor components) could represent other items or relations\eg the significance of accumulated events in a monitoring task, the weight of edges in a graph, or the weight of word-context relations in natural language processing.
Each component of the tensor is first populated with a random integer drawn from the flat distribution $[0,10]$.
This represents noise and/or insignificant features of the classes.
We then add a relatively low number of significant features to each class, which are represented by high-value components.
The value of these components, $w$, is taken to be 100 or 1,000.
This can be translated into a signal-to-noise ratio (SNR).
The SNR can be expressed in the root-mean-square (RMS) amplitudes of signal and noise, $\text{SNR} = \left(A_{\text{signal}}/A_{\text{noise}}\right)^2$.
We take the number of features to be proportional to the size of the matrix, $N_{\cal D}$, and define the constant of proportionality as $\rho$.
The number of control features in each class then is $\rho N_{\cal D}$, each having weight $w$.
The RMS amplitude of the signal is $\sqrt{\rho N_{\cal D} w^2 / N_{\cal D}} = \sqrt{\rho}w$.
The RMS amplitude of a uniform random distribution on the interval $[0,M]$ follows from the definition and is $\sqrt{M(2M+1)/6}$.
It then follows that the SNR can be expressed in $\rho$, $w$ and $M$ in the following way
\bea
	\text{SNR} &=& 10~\text{log}\left(\frac{A_{\text{signal}}}{A_{\text{noise}}}\right)^2~\text{dB} \nonumber \\
	 &=& 10~\text{log}\left(\frac{6 \rho w^2}{M(2M+1)}\right)~\text{dB}.
	 \label{eq:snr}
\eea
For example, the SNR would be $5$~dB if the background is uniformly distributed on the interval $[0,10]$ and there are one percent features, $\rho=0.01$, with weight $w=100$.
The SNR of decoded features is lower, because the RI representation is approximate.
Note that a high SNR does not automatically imply that the data can be accurately represented with NRI\eg because a high value of $\rho$ can make the representation inaccurate, see the discussion above.

\begin{figure*}[h]
\centering
\begin{minipage}[b]{7.5cm}
\centering
\includegraphics[width=\linewidth]{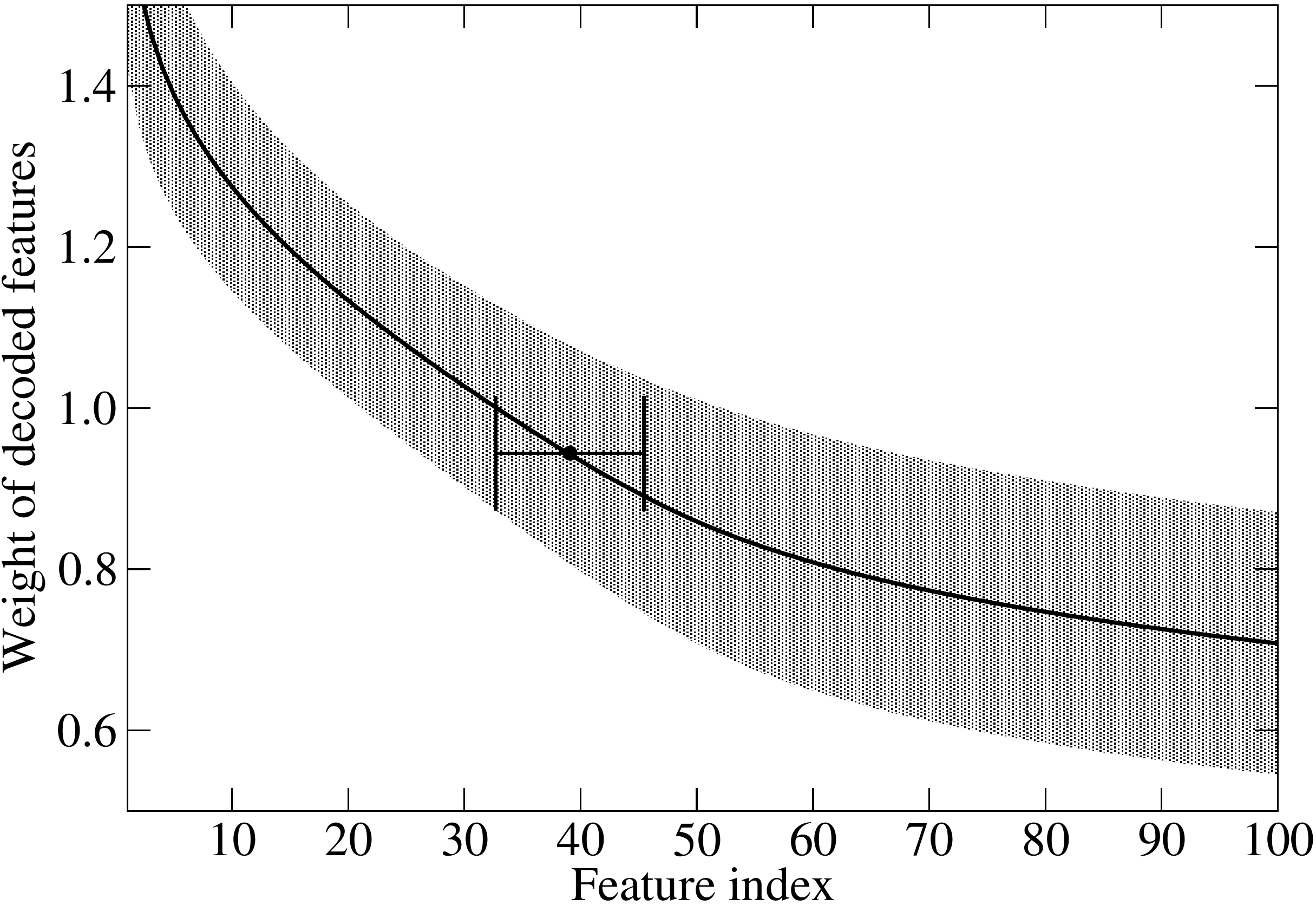}
\end{minipage}
\hspace{0.5cm}
\begin{minipage}[b]{7.5cm}
\centering
\includegraphics[width=\linewidth]{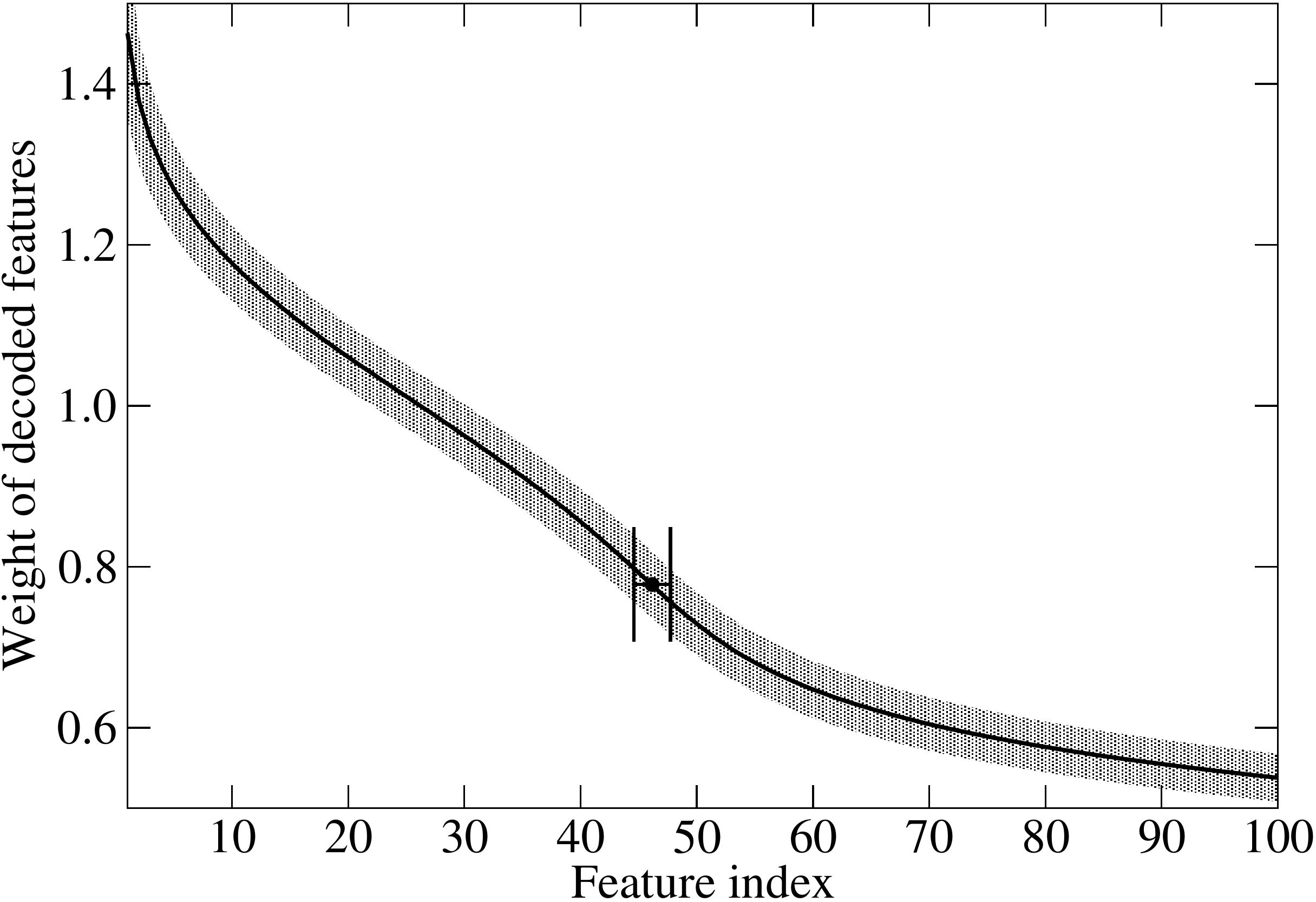}
\end{minipage}
\caption{Average decoded component weights of a rank-two tensor encoded with two-way RI.
In this example a $10,000\times 10,000$ matrix is encoded in a $5,000\times 5,000$ state\ie the dimension reduction is 4:1.
Each column of the matrix represents a ``class'' and each row represents a ``feature'' that the classes may have, see the text for an explanation.
The matrix is initialized with uniform random numbers in the range $[0,10]$.
Fifty randomly selected features are then added to each of the ten thousand classes.
In the figure on the (left-) right-hand side these features have weight (100) 1,000, which corresponds to a signal to noise ratio of about (1.5 dB) 21 dB.
The solid curves illustrate the average decoded weight, which has been normalized with the weight of control features (100 and 1,000, respectively).
The horizontal axis is a counter of the first one hundred top-score components returned by the decoding function.
The shaded area represents the standard deviation of the weights obtained for the ten thousand different classes.
The data points with horizontal error bars represent the average and standard deviation of the number of correctly identified features within the first fifty top-score elements.
With a weight of 100 (1,000) of the control features the number of correct features out of 50 is $39 \pm 6.4$ ($46 \pm 1.6$). 
}
\label{fig:toplist}
\end{figure*}

\begin{figure}[h]
\centering
\includegraphics[width=\figurewidth]{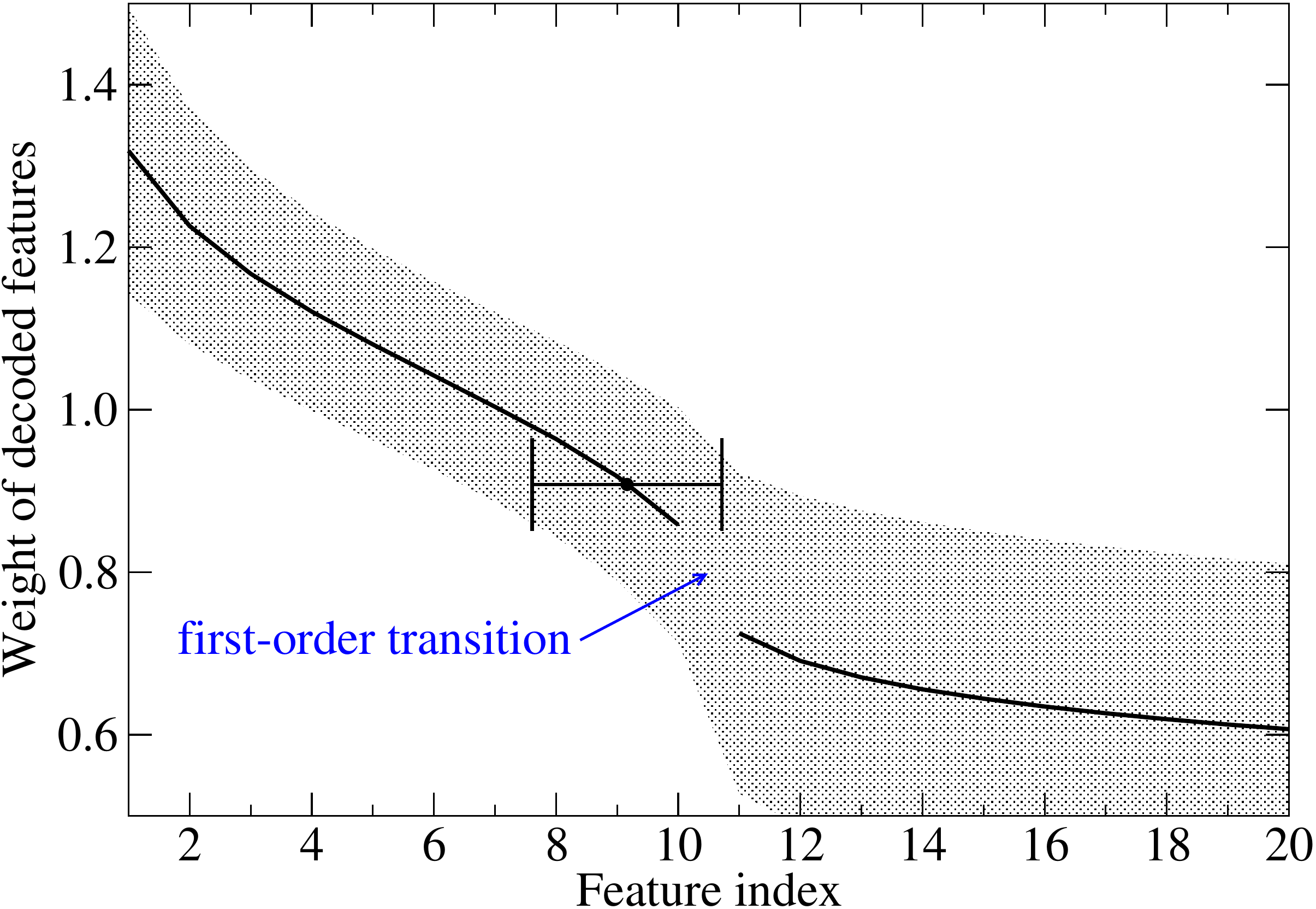}
\caption{
Example of a first-order transition in the average decoded weights of a rank-two tensor encoded with two-way RI.
The case illustrated here is identical to the left-hand side panel of \fig{toplist} with the only difference that the number of control features for each class is ten instead of fifty.
In this case $9.2 \pm 1.6$ features out of ten are identified via the top-ten list of decoded high-value matrix elements.
This is significantly better than the result presented in \fig{toplist} and the transition between features and noise is distinct.
}
\label{fig:toplist_fo}
\end{figure}

In the first example we consider a matrix of size $10,000 \times 10,000$ that is encoded with two-way RI in a $5,000\times 5,000$ state\ie the dimension reduction is 4:1.
This implies that the index vectors have length $n_{\cal D} = 5,000$.
Unless stated otherwise we will use index vectors with four positive and four negative trits so that $\chi_{\cal D}=8$.
The motivation for this choice is given below.
For each of the $10,000$ classes we add $50$ randomly selected features, which are referred to also as control features.
These features are given a weight of $100$ in one case, and $1,000$ in another case.
The question then is, how many of these fifty control features can be identified?
To answer this question we extract the fifty matrix elements of each class that have the highest decoded weights and compare these with the features that actually were encoded in the matrix.
The result of that simulation is presented in \fig{toplist}, which includes the decoded weights of the first one hundred elements.
The answer to the question posed above is that $39 \pm 6.4$ ($46 \pm 1.6$) of the fifty control features are represented among the top-fifty matrix elements of each class when the weight of the features is $100$ ($1,000$).
These numbers are somewhat higher if the matrix is encoded with one-way RI and the standard deviation of the decoded weights is lower in that case also, we will return to that below.
A number of things can be learned from the result presented in \fig{toplist}:
1) for a fixed number of features and a fixed weight of the features, a higher weight tends to increase the probability that the features are correctly identified via the decoded value of the matrix elements,
2) the standard deviations of the number of correctly identified features and the decoded weights are lower when the weight is higher,
3) it appears that the transition from control features to random noise at an index of 50 is more distinct with the higher weight of $1,000$,
and 4) the decoded weights of the top-fifty elements is distributed around the encoded weight and can deviate from it by at least $\pm 50$ percent.
The transition from features to noise in the top-list of decoded weights is in fact distinct in some cases and then resembles a first-order phase transition in thermodynamic systems.
This point is illustrated in \fig{toplist_fo}, which corresponds to the left-hand side panel of \fig{toplist} with the only modification being that $\rho=0.001$\ie there is only $0.1$ percent features in each class.
A formal analysis of the fidelity of encoded tensor components is beyond the scope of this work.
The interested reader is referred to Kanervas book \cite{SDM_1988} for a similar discussion.
This is an interesting issue for further investigation, because it is related to the question of how to select the length of the list of decoded high-value elements when the number of significant features in the classes is unknown.
Our results provide two guiding principles.
First, the reliability of the high-value elements tend to decrease with their index in the descending list.
Second, when the relative number of features is low and the relative weight of features is high the decoded weights of features is separated from the lower weights of non-features in a distinct transition.

\begin{figure*}[h]
\centering
\begin{minipage}[b]{7.5cm}
\centering
\includegraphics[width=7.5cm]{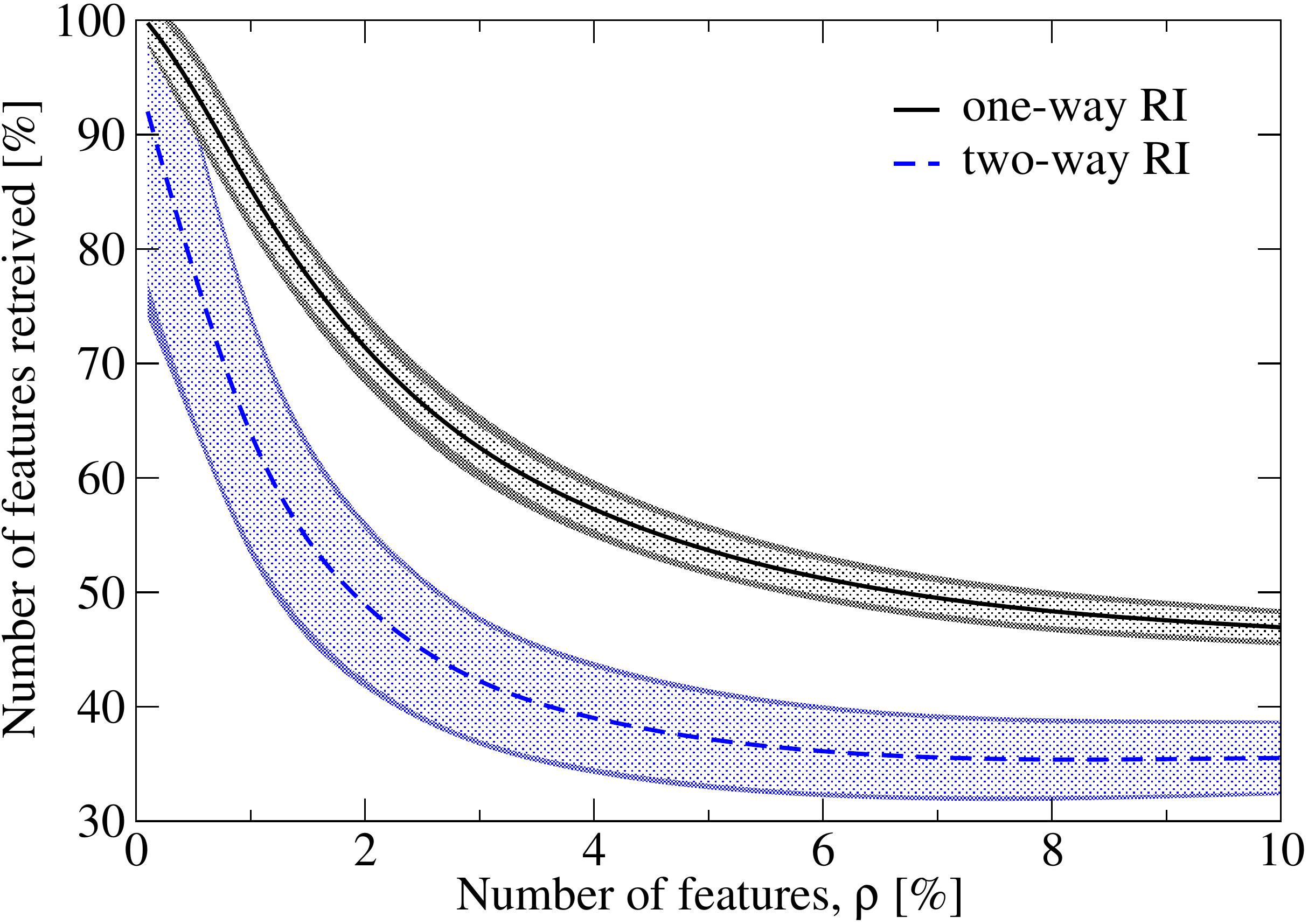}
\end{minipage}
\hspace{0.5cm}
\begin{minipage}[b]{7.5cm}
\centering
\includegraphics[width=7.5cm]{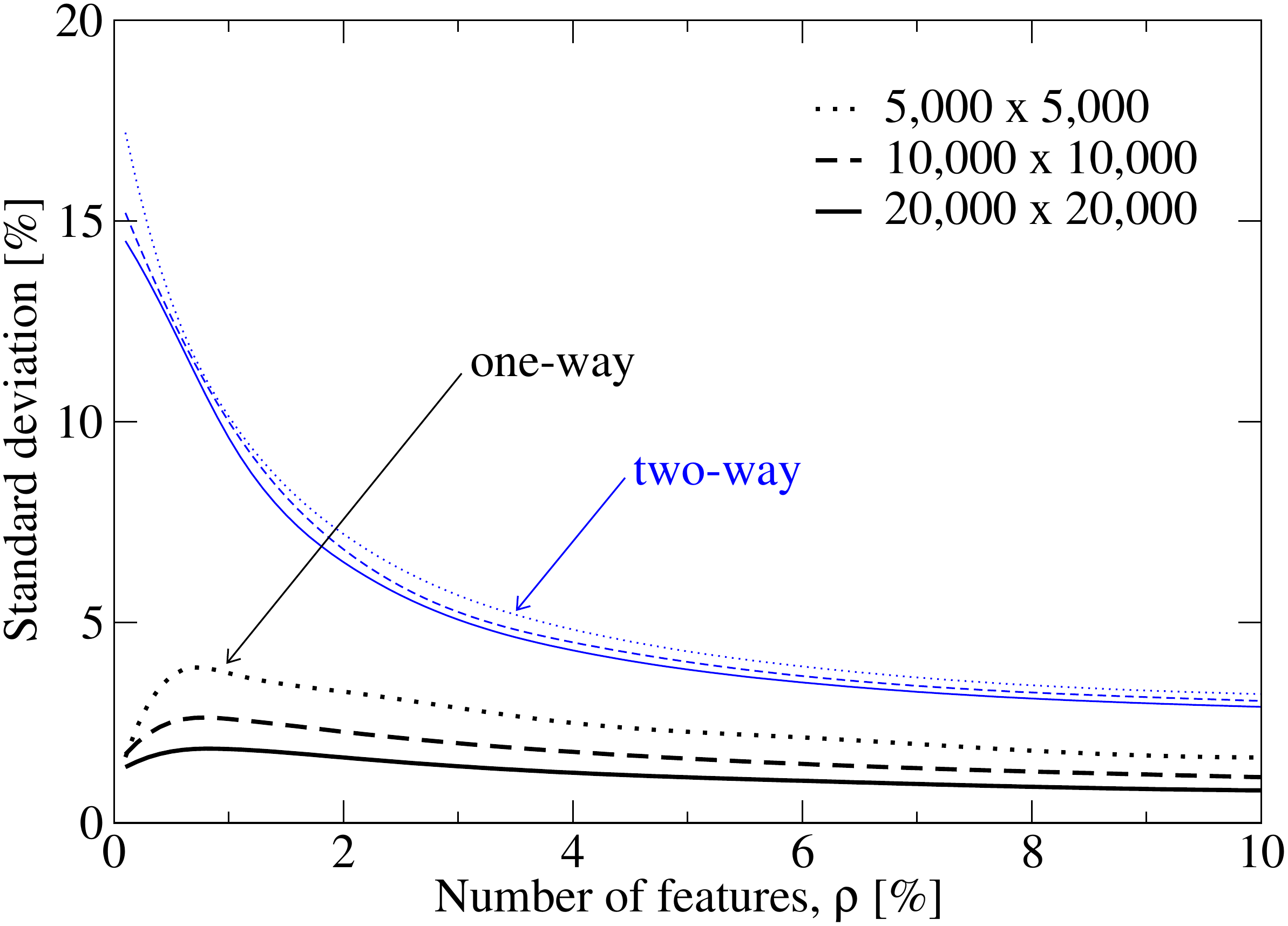}
\end{minipage}
\caption{
The number of correctly decoded features represented in a rank-two tensor encoded with one-way and two-way RI, see \fig{toplist} and the text for further information.
The horizontal axis represents the relative number of encoded control features, $\rho$, and the vertical axis of the panel on the (right-) left-hand side represents the average (standard deviation) of the relative number of correctly decoded features.
The average is practically independent of the size of the matrix and the dimensionality of the index vectors, provided that the dimension reduction ratio is kept constant, which is the case here.
The standard deviation of the number of correctly decoded features does, however, decrease with increasing dimensionality of the index vectors.
The shaded areas in the panel on the left-hand side illustrate the standard deviations for two different dimensionalities of the index vectors.
In the case of one-way (two-way) RI the higher standard deviation corresponds to a $5,000 \times 5,000$ matrix encoded in $1,250 \times 5,000$ ($2,500 \times 2,500$) states, while the lower standard deviation corresponds to a $10,000 \times 10,000$ matrix encoded in $2,500 \times 10,000$ ($5,000 \times 5,000$) states.
The results presented in the panel on the right-hand side corresponds to these dimensionalities also, and it includes an additional result for a $20,000 \times 20,000$ matrix that is encoded with the same dimension reduction of 4:1.
The higher dimensionality of the index vectors in the $20,000 \times 20,000$ case yields a lower standard deviation than obtained in the other two cases.
}
\label{fig:onevstwo}
\end{figure*}

Next we turn to a comparison of one-way and two-way RI, and a study of the effects of varying the dimensionality of index vectors.
The philosophy here is identical to that above\ie we consider matrices that encode features of classes.
We vary the relative number of features, $\rho$, from $0.1$ to $10$ percent of the size of the matrix, $N_{\cal D}$.
If $\rho N_{\cal D}$ control features are encoded in a class then a top-list of equally many high-value matrix elements is decoded and the number of control features in that top-list determines the relative number of features retrieved.
This methodology is the same as that used in the analysis leading to the data points with error bars in \fig{toplist}.
The dimension reduction, $\Pi_{\cal D} N_{\cal D}$~:~$\Pi_{\cal D} n_{\cal D}$, is kept fixed at 4:1 in this example also, but the size of the matrix is varied so that the dimensionality of the index vectors varies.
From the analysis of orthogonality in \sect{ternary} one could expect that the relative number of correctly decoded features should increase with dimensionality.
It turns out that this is not the case.
Instead we find that the relative number of correctly decoded features is practically {\em independent} of dimensionality, but that the standard deviation decreases with increasing dimensionality.
In other words, when the dimension reduction ratio is kept fixed and the number of encoded features is proportional to the size of the matrix the effect of increasing the size of the matrix and thereby the dimensionality of index vectors is to reduce the uncertainty in the number of correctly decoded features.
This point is illustrated in \fig{onevstwo}.
A prerequisite of this conclusion is that the dimensionality is sufficiently high\ie of the order of a thousand components.
It is not valid for low-dimensional index vectors, $n_{\cal D} \lesssim 1000$, because in that domain the average performance decreases.
The details of this performance degradation are presently not understood.
Note that the relative number of correctly decoded features first decreases with an increasing number of encoded features, as expected, but then starts to increase somewhat for $\gtrsim 8$ percent features in the case of two-way RI.
This effect is caused by the increasing probability of obtaining features in the top-list by chance when the relative number of features and the relative length of the top-list increases.
In the case of one-way RI the standard deviation has a maximum around $0.7$--$0.9$ percent features and the nature of this behaviour is not understood. 
It may be an effect of chance similar to that described above, but the low value of $\rho$ at the maxima renders such an explanation counter intuitive.
We see no major reason to explore this property in more depth here and therefore leave it as an open issue.

\begin{figure*}[h]
\centering
\begin{minipage}[b]{7.5cm}
\centering
\includegraphics[width=7.5cm]{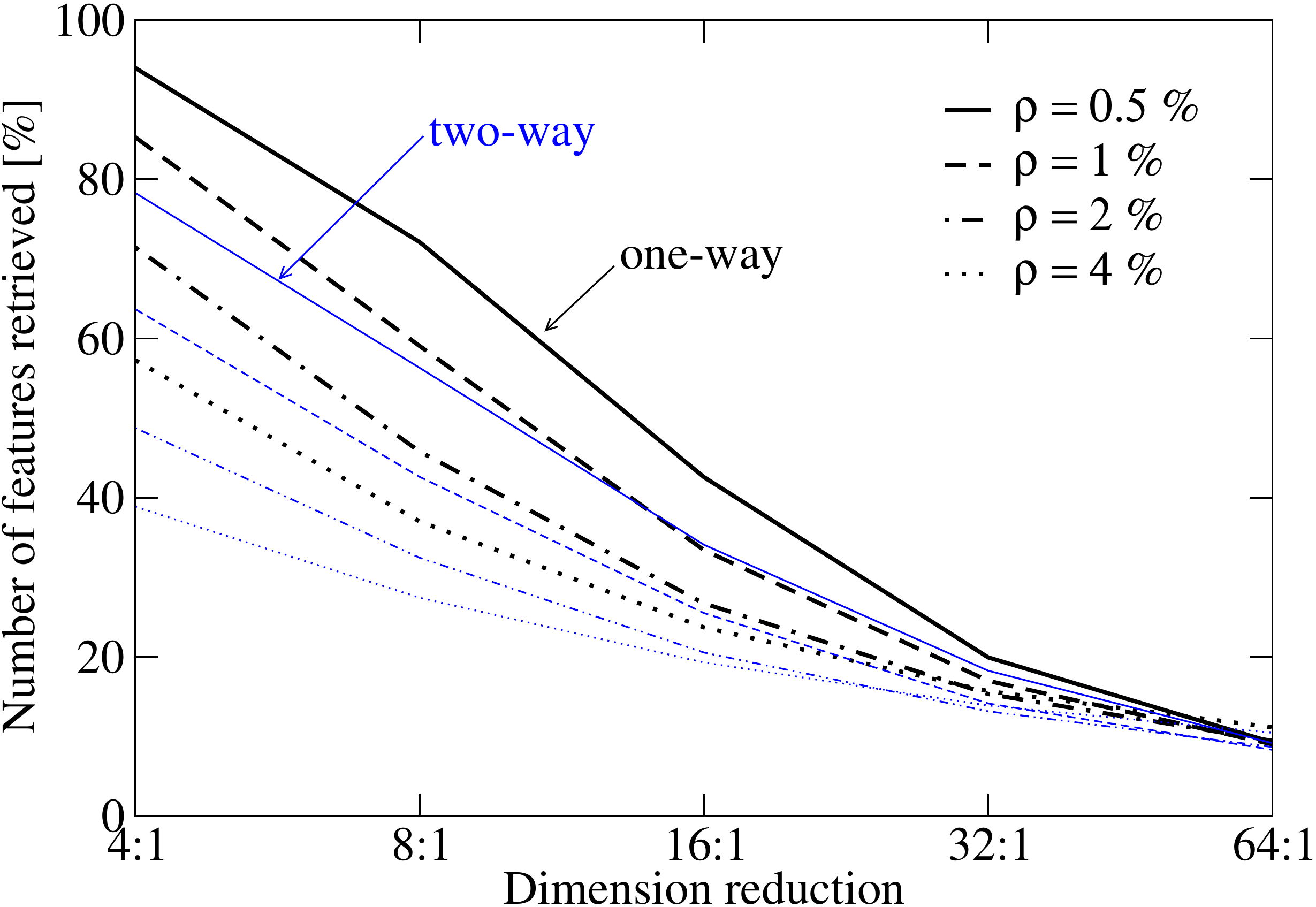}
\end{minipage}
\hspace{0.5cm}
\begin{minipage}[b]{7.5cm}
\centering
\includegraphics[width=7.5cm]{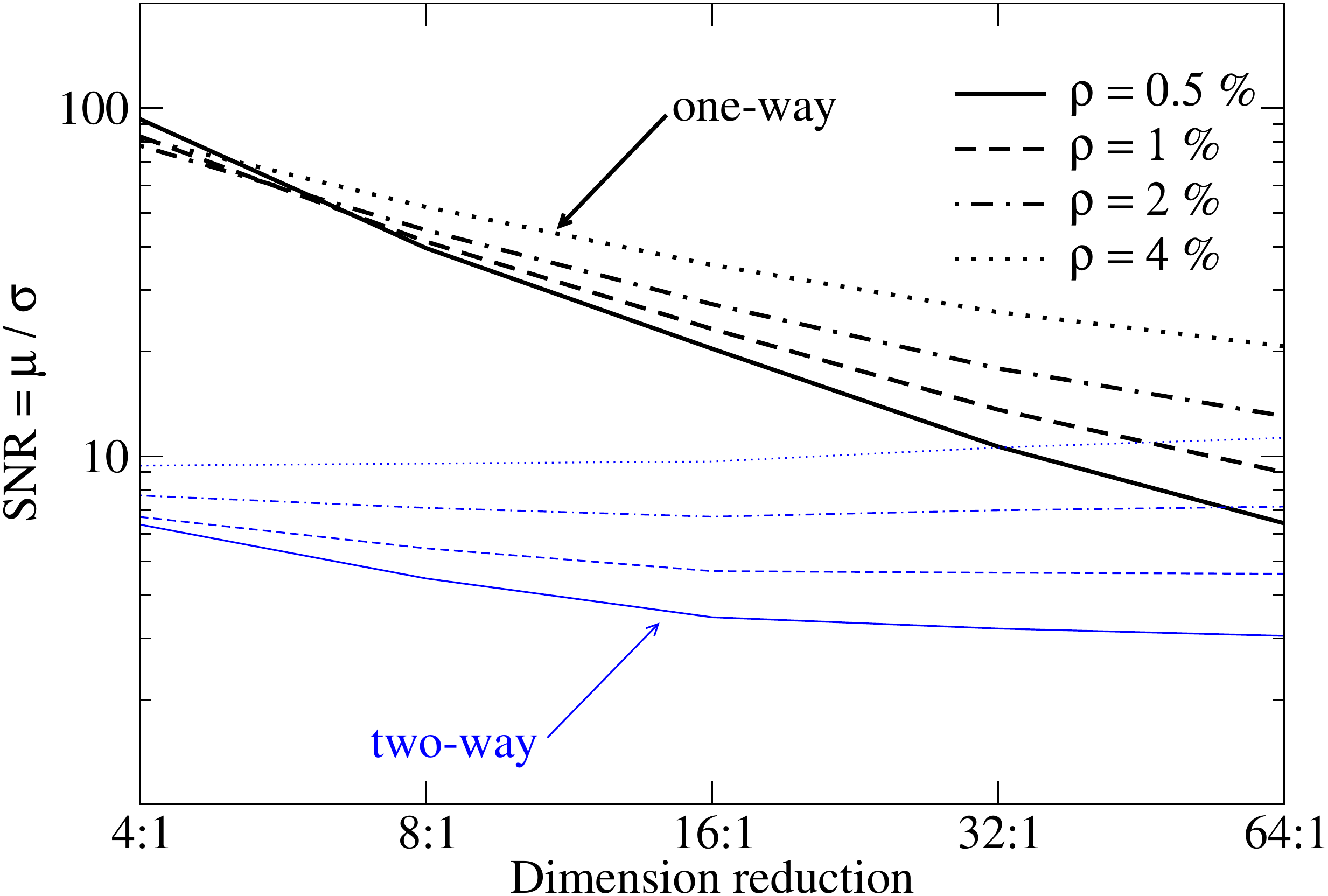}
\end{minipage}
\caption{
Effect of the dimension reduction, $\Pi_{\cal D} N_{\cal D}$~:~$\Pi_{\cal D} n_{\cal D}$, on the relative number of correctly decoded features.
The panel on the left-hand side shows the average relative number of correctly decoded features, which is practically independent of matrix size as long as the dimensionality of the index vectors is sufficiently high, see the text for further information.
The panel on the right-hand side shows the signal to noise ratio, defined as the average relative number of correctly decoded features, $\mu$, divided by the corresponding standard deviation, $\sigma$.
Note that this signal to noise ratio refers to the decoded information, while \eqn{snr} refers to the encoded data.
The size of the matrix, $N_{\cal D}$, is taken to be $64,000 \times 64,000$ for both one-way and two-way RI.
At the maximum dimension reduction of 64:1 this corresponds to a state size, $n_{\cal D}$ of $1,000 \times 64,000$ ($8,000 \times 8,000$) for one-way (two-way) RI.
The large size of the matrix is necessary to maintain high dimensionality of the index vectors for one-way RI at high dimension reduction.
The effect of an increasing dimension reduction ratio on the dimensionality of index vectors is lower for two-way RI than for one-way RI, and is increasingly weaker at higher order.
}
\label{fig:dimred}
\end{figure*}

Up to this point the dimension reduction is kept fixed at 4:1.
The next question is how the dimension reduction ratio affects the performance of NRI.
We address this question with an example that is identical to those considered above with the only modification that the dimension reduction, $\Pi_{\cal D} N_{\cal D}$~:~$\Pi_{\cal D} n_{\cal D}$, is varied from 4:1 to 64:1.
The size of the matrix, $N_{\cal D}$, is kept constant when varying the dimension reduction ratio, which implies that the number of features encoded in the classes is constant also.
The result of this simulation is presented in \fig{dimred}.
A surprising result of this simulation is that the performance of one-way and two-way RI is comparable at a dimension reduction of 64:1.
The nature of this trend may be related to an interesting scaling phenomena.
Assume that a matrix is square and of size $N \times N$, and that it is encoded with one-way (two-way) RI in a state of size $n \times N$ ($n \times n$).
The dimension reduction ratio, $\xi$, then is $\xi = N/n$ ($\xi = N^2/n^2$) for one-way (two-way) RI.
Solving for the dimensionality of the state matrix, and consequently of the index vectors, we get $n = N/\xi$ for one-way RI and $n = N/\sqrt{\xi}$ for two-way RI.
In general, for any order, $r$, of NRI the dimensionality of the index vectors scales as $n = N \xi^{-1/r}$.
The order of NRI ($r$) therefore effectively reduces the impact of high dimension reduction ratios on the dimensionality of index vectors.
Due to the high computational cost of systematic simulations at high dimension reduction ratios and tensor rank we do not investigate these trends further here, but leave them as an interesting issue for further investigation.

\begin{figure*}[h]
\centering
\begin{minipage}[b]{7.5cm}
\centering
\includegraphics[width=7.5cm]{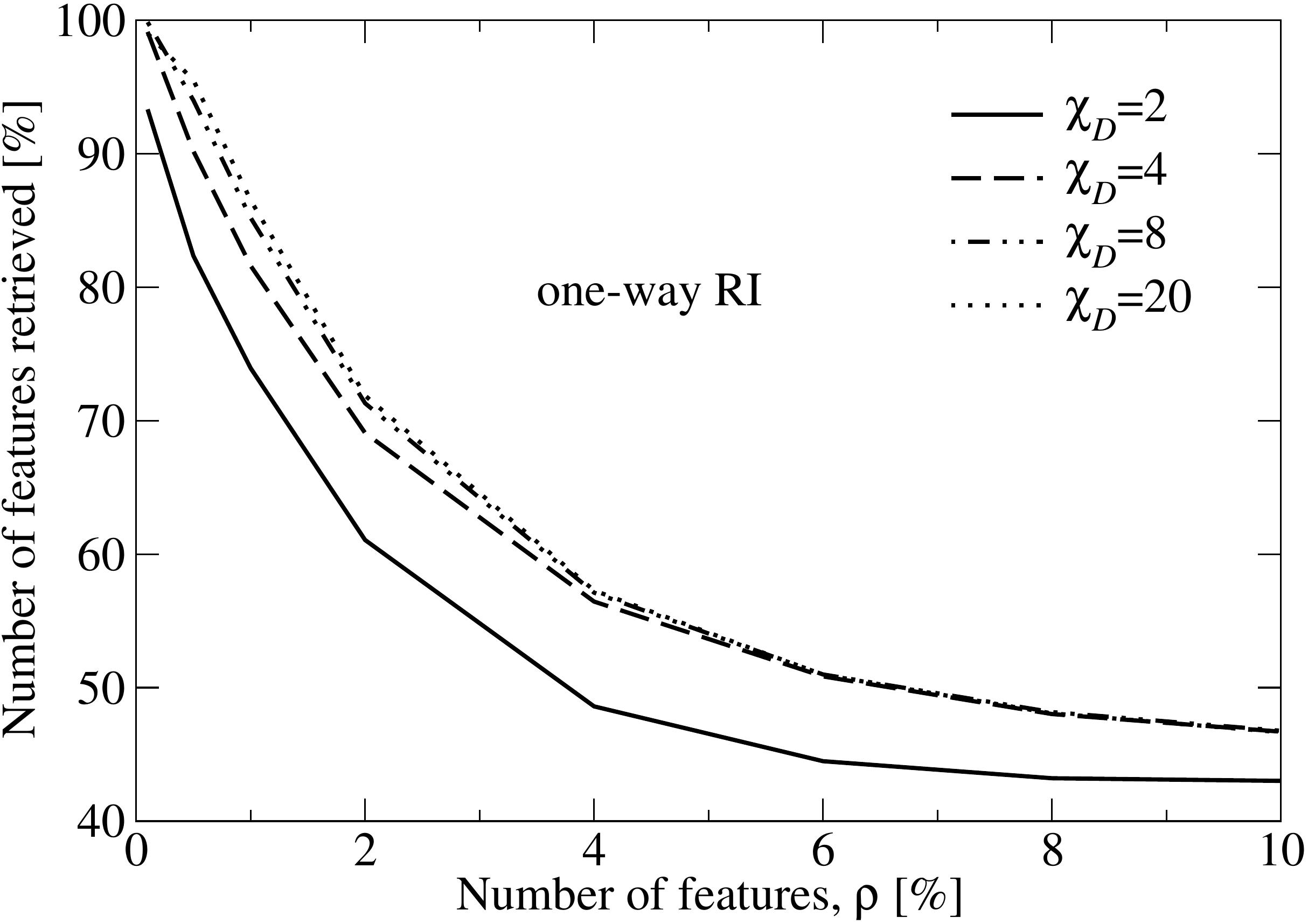}
\end{minipage}
\hspace{0.5cm}
\begin{minipage}[b]{7.5cm}
\centering
\includegraphics[width=7.5cm]{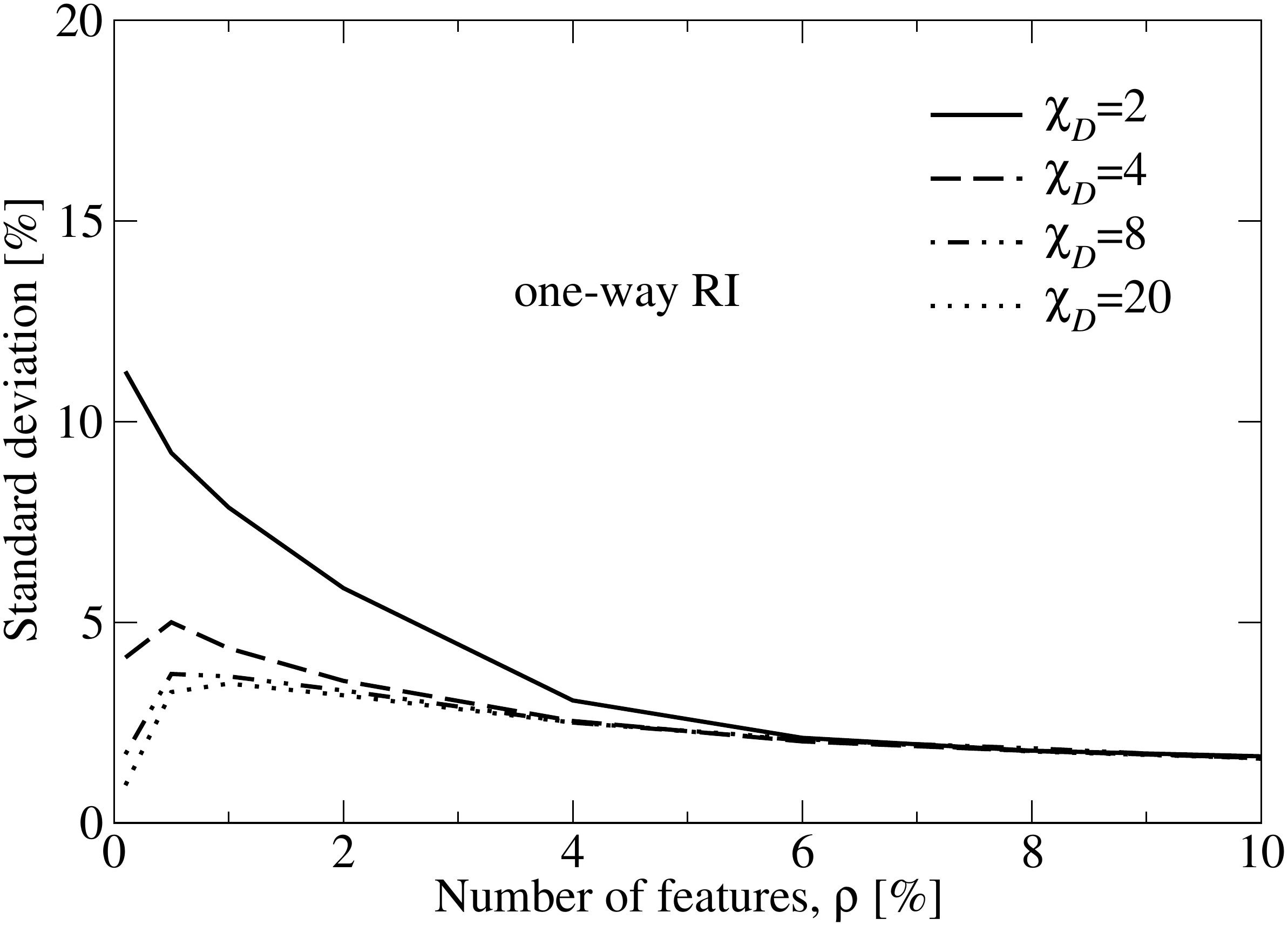}
\end{minipage}
\vspace{0.3cm} \\
\begin{minipage}[b]{7.5cm}
\centering
\includegraphics[width=7.5cm]{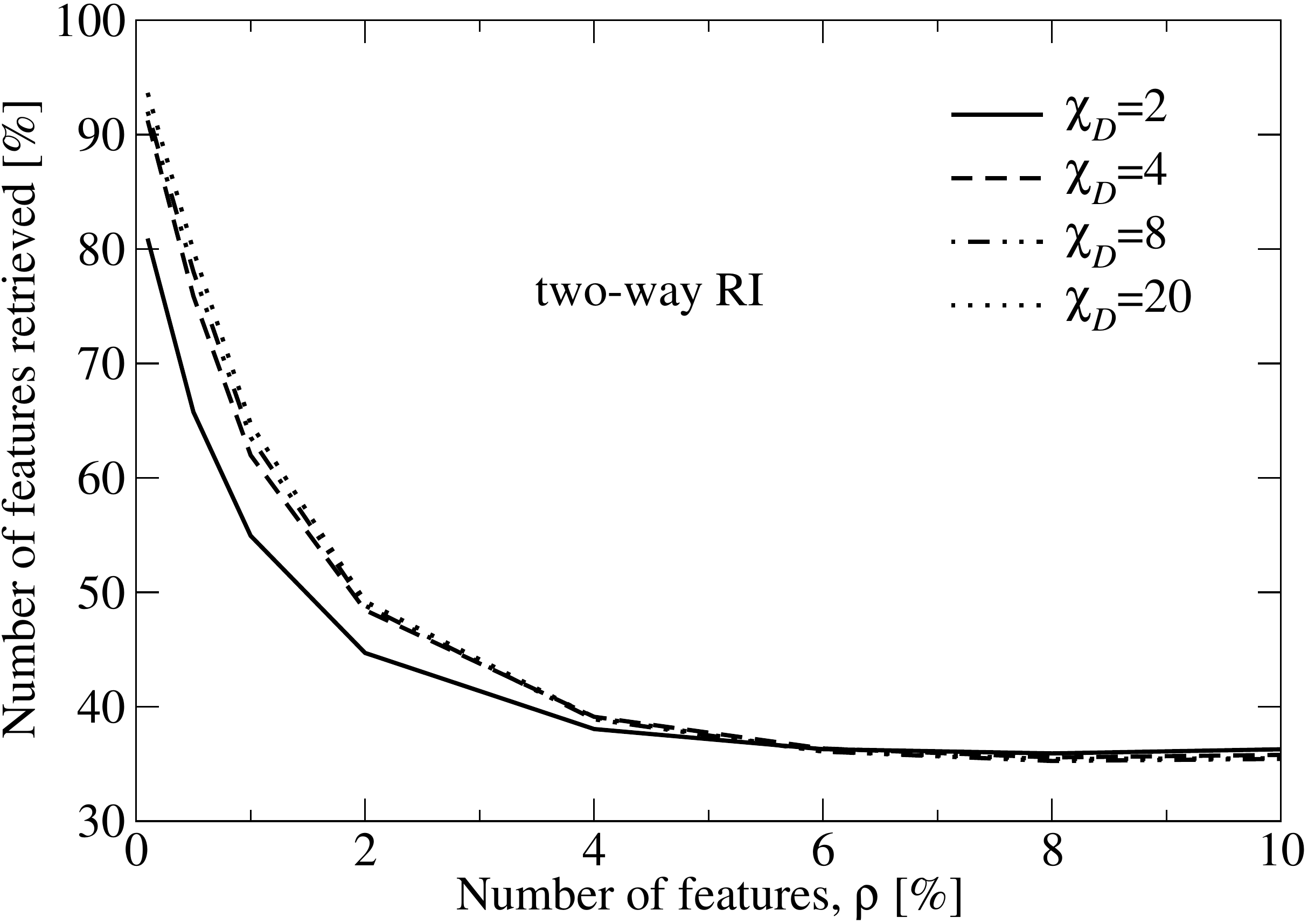}
\end{minipage}
\hspace{0.5cm}
\begin{minipage}[b]{7.5cm}
\centering
\includegraphics[width=7.5cm]{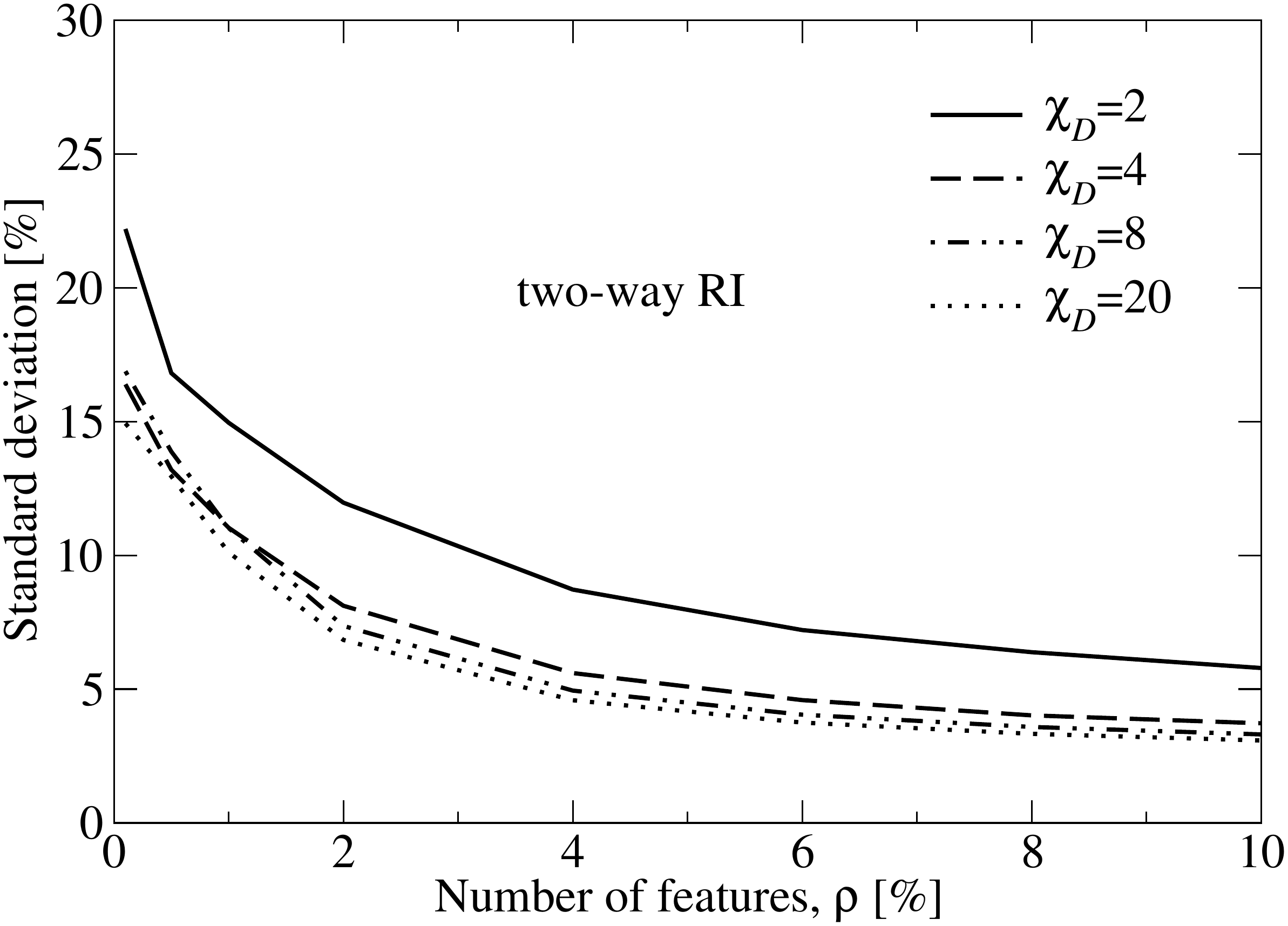}
\end{minipage}
\caption{
The average number of correctly decoded features and the corresponding standard deviation for different numbers of non-zero trits in the index vectors, $\chi_{\cal D}$, and different relative number of encoded features, $\rho \in \{0.1,~0.5,~1,~2,~4,~6,~8,~10\}$ percent.
The matrix considered here has size $5000 \times 5000$ and it is encoded with one-way and two-way RI such that the dimension reduction is 4:1.
Results for $\chi_{\cal D}=8$ are identical to those presented in \fig{onevstwo}.
Refer to that figure and the related text for further information about the $\rho$-dependence of the quantities plotted here.
}
\label{fig:diffk}
\end{figure*}

Finally we investigate the dependence of these results on the number of non-zero trits, $\chi_{\cal D}$, in the index vectors.
Up to this point we kept this parameter fixed at $\chi_{\cal D} = 8$\ie we used index vectors with four positive and four negative trits.
This is a good compromise, because the performance in the examples considered here increases insignificantly at higher values of $\chi_{\cal D}$, and the computational complexity increases with increasing $\chi_{\cal D}$ since the number of states that needs to be accessed each time a tensor component is encoded or decoded is $\Pi_{\cal D}\chi_{\cal D}$, see \sect{ricoding}.
In \fig{diffk} we illustrate how the average relative number of correctly decoded features varies for different values of $\chi_{\cal D}$ and the relative number of encoded features, $\rho$.
Illustrated in that figure is also the corresponding standard deviation.
From these results it is clear that there is a significant improvement in performance when increasing $\chi_{\cal D}$ from two to four, and that the improvement thereafter is relatively small.
In two-way RI the number of states associated with one matrix element is $\chi_{1} \times \chi_{2}$ so there is practically a quadratic dependence of the computational complexity on the number of non-zero trits in the index vectors, $\chi_{\cal D}$.
This is the motivation why we settled for $\chi_{\cal D}=8$ in most simulations presented in this section, since it offers practically the same performance as higher values, but at a lower computational cost.
Note the maxima of the standard deviation around $\rho\sim 0.7$ percent in the case of one-way RI, which we noted and commented on also in relation to \fig{onevstwo}.
Here an exception occurs for $\chi_{\cal D}=2$, which does not have such a maxima.
Another surprising feature of these results is the intersection of standard deviations for two-way RI around one percent encoded features.
For low $\rho$ the standard deviation is marginally higher with $\chi_{\cal D}=8$ than $\chi_{\cal D}=4$.
The nature of this detail is not understood.

%%%%%%%%%%%%%%%%%%%%%%%%%%%%%%%%%%%%%%%%%%%%%%%%%%%%%%%%%%%%%%%

\section{Application to natural language analysis}
\label{sec:language}

Next we apply this methodology in a statistical study of natural language to further illustrate the NRI approach and some of its properties.
In statistical models of natural language it is common to construct a large word--word or word--document matrix, a so-called {\it co-occurrence} matrix.
This is for example the case in the Hyperspace Analogue to Language (HAL) \cite{lund_1995,lund_1996} and in Latent Semantic Analysis (LSA) \cite{landauer_1997}, which are two pioneering and successful models in this field.
In practical applications the number of words normally exceeds hundreds of thousands.
The number of contexts is necessarily high also, otherwise the statistical information will be insufficient for successful analysis.
Word co-occurrence matrices therefore tend to be large objects\eg the relatively simple example considered in the following uses more than 5 billion matrix elements. 
Fortunately, co-occurrence matrices are sparse and can therefore be compressed to make the semantic analysis more efficient.
It was demonstrated in \cite{kanerva_2000} that one-way RI can be used to effectively encode co-occurrence matrices for semantic analysis, see also \cite{sahlgren_2005,sahlgren_thesis,kanerva_hyperdimensional_2009}.

The definition of context is model specific, but typically includes a set of words or a document.
In HAL the context is defined by a number of words that immediately surround a given word, while in LSA the context is defined as the document where the word exists.
Linguistically, the former relation can be described as a paradigmatic\ie semantic relation, while the latter can be characterized as an associative\ie topical relation.
In the traditional RI algorithm, each word type that appears in the data is associated with a context vector, and each context is associated with a ternary index vector.
If the context is defined in terms of the neighbouring words of a given word, which is the strategy that we adopt here, context vectors are created by adding the index vectors of the nearest (typically two or three) preceding and succeeding words every time a word occurs in the data.
If the context is defined as the document where the word exists, context vectors are created by adding the index vectors of all documents where a word occurs, weighted by the frequency of the word in each document.
In either case a context vector is the sum of weighted index vectors of all contexts, defined either by surrounding words or by documents, where that word occurs.
The RI algorithm has traditionally been evaluated using various kinds of vocabulary tests\eg the synonymy part of the ``Test of English as a Foreign Language'' (TOEFL)
\cite{kanerva_2000,sahlgren_thesis}.

In the following we revise the synonym identification task presented in \cite{kanerva_2000} with three changes.
First, we want to compare the performance of one-way and two-way RI, so we encode the co-occurrence matrix with both methods.
Second, while Kanerva \etal used the LSA definition of context, we use a strategy similar to that in HAL and define the context as a window that spans $\pm 2$ words away from the word itself.
This means that for each occurrence of a word there will be four additional word--word correlations encoded in the co-occurrence matrix.
Phrased differently, our co-occurrence matrix is not a word--document matrix, but a word--word matrix where context is defined in terms of word neighbourhood.
This strategy avoids the potential difficulty of defining document (context) boundaries\eg in on-line text, and it captures semantic relations between words rather than topical relations.
The length of the context window is a free parameter that affects the quantitative results presented here, but it is not essential for our qualitative discussion.
The third and last difference compared to the study in \cite{kanerva_2000} is that we do not introduce upper and lower limits for the frequencies encoded in the co-occurrence matrix.
High-frequency word-context relations have a negative effect on NRI representations, because any interference (non-orthogonality) between index vectors is amplified by the high weight of the most frequent correlations.
Frequently occurring words like ``the'', ``at'' and ``be'' contribute to extremely high frequencies that render the occurrences of more interesting combinations insignificant, practically by pushing them below the noise threshold.
This effect is stronger for two-way RI than for one-way RI, because in one-way RI the interference affects one dimension only\eg contexts interfere but words are kept independent.
In two-way RI both words and contexts interfere and high-frequency word-context combinations needs to be transformed or removed prior to encoding.
The significance of this effect presumably increases with tensor rank, but we have not made simulations to investigate that trend.
Here we include the complete word-context spectrum, including the low- and high-frequency content, and we present results for two different transformations of the spectrum.
In one case we encode the unaltered frequencies directly, and in the other case we take the square root of the frequencies before encoding them.
The effect of the square root is to decrease the significance of high frequencies, and this has a positive effect on the SNR of the RI-encoded co-occurrence matrix.
This choice of transformation was inspired by the qualitative relation $E\sim A^2$ for wave phenomena (where $E$ denotes energy and $A$ amplitude), but it has no obvious connection to that relation and is adopted here without further motivation.
We made some experiments with a logarithmic rescaling of frequencies also, but that seems to be inferior to the square root.
No systematic comparison of transformations was made, and no further efforts were made to optimize the transformation for maximum performance since our focus is on qualitative differences between one-way and two-way RI.
The quantitative results can be further improved with a more careful choice of preprocessing method\eg by introducing cuts on the word--word correlation frequencies \cite{karlgren_2001}.
Another example is the preprocessing method used in LSA, where the frequencies are transformed with a logarithm and then divided with the conditional entropy of the context given that the word has occurred \cite{landauer_1997}.
Note, however, that such preprocessing methods invalidate the incremental property of NRI, because the accumulated frequency (weight) must be known before the transformation is made.
This problem can possibly be addressed with a decoding-encoding step when incrementing the matrix elements and/or a more sophisticated pre-processing technique.

\begin{figure}[h]
\bigskip
\includegraphics[width=\figurewidth]{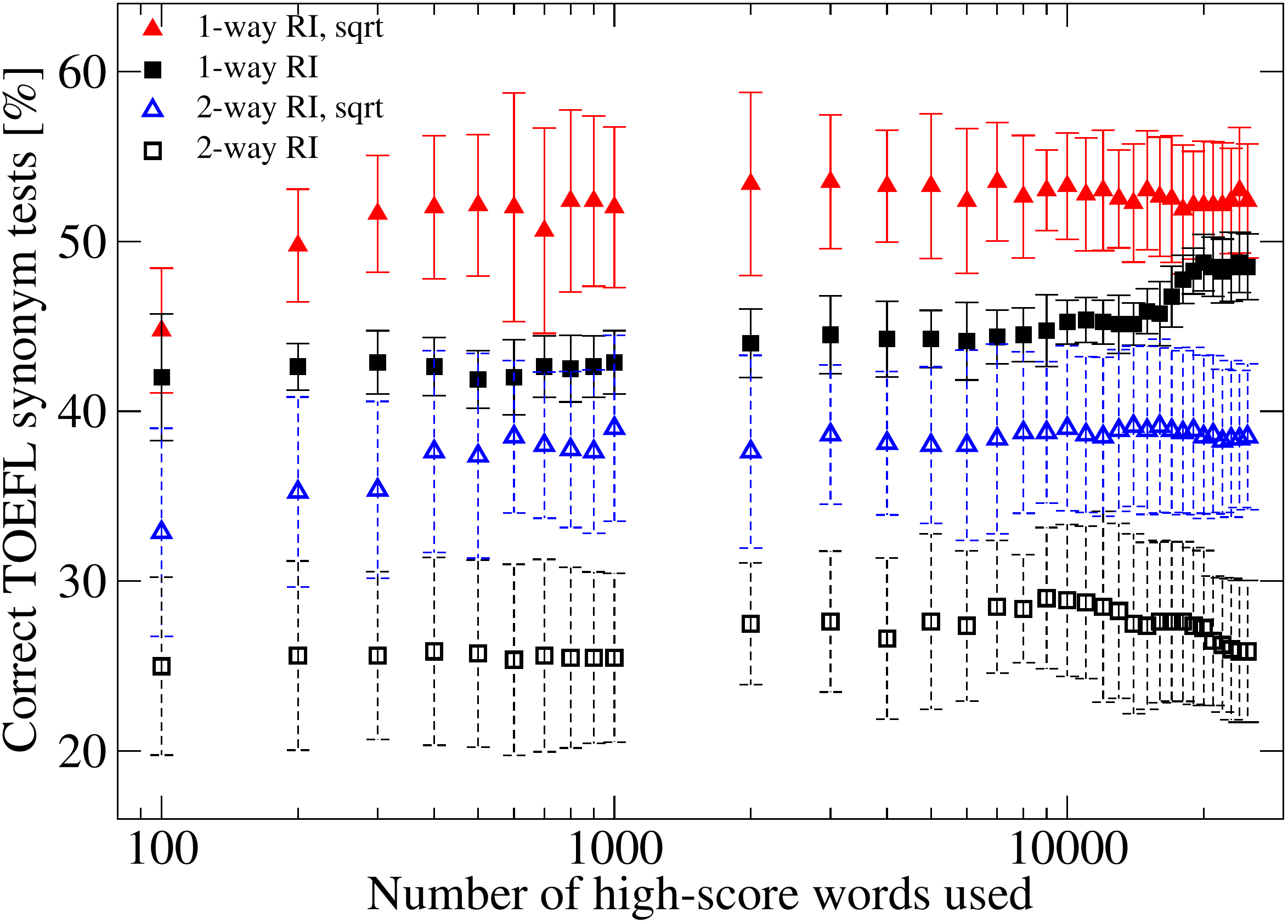}
\caption{
Performance of different NRI techniques on TOEFL (Test Of English as a Foreign Language) synonym tests.  
Illustrated here is the number of correctly identified synonyms vs. the size of the word--word correlation sets used to calculate the Jaccard index \eqn{jaccard}.
The co-occurrence matrix is encoded with NRI and includes 37,620 short high-school level articles in the TASA (Touchstone Applied Science Associates, Inc.) corpus.
Each TOEFL synonym test consists of five words, out of which two are synonyms.
The task is to identify the synonym of a given word in the set\ie there are four alternative answers out of which one is correct.
This implies that an average score of $25$ percent corresponds to random guesses.
These results are based on 80 synonym tests, each comprising five words.
Each test were repeated ten times with new randomly generated index vectors.
Symbols represent average results of the ten simulations, and error bars represent standard deviations.
When performing the synonym similarity test in the traditional way using one-way RI and the cosine of angles between full-length context vectors the result is $47\pm 2.4$ percent correctly identified synonyms, which increases to $51\pm 3.7$ percent when the square root of word frequencies is used.
This is slightly lower than the best results obtained with the Jaccard index test, which are $49\pm 1.7$ and $54\pm 3.9$ percent, respectively.
See the text for further information.
}
\label{fig:toefl}
\end{figure}

We construct the co-occurrence matrix from 37,620 short high-school level articles in the TASA (Touchstone Applied Science Associates, Inc.) corpus
\footnote{The TASA and TOEFL items have been kindly provided by Professor Thomas Landauer, University of Colorado.}.
The text has been morphologically normalized so that each word appears in base form.
It contains about 74,200 word types that are encoded in an co-occurrence matrix with the RITensor software \cite{RITensor}.
For one-way RI we use index vectors of length $1,000$, so that the dimension reduction is $\sim 74,200\times 74,200~:~1,000\times 74,200 \rightarrow 74~:~1$.
In the case of two-way RI we use a state tensor of size $~1,000\times 74,200$, thereby maintaining the same dimension reduction ratio.
We have repeated the two-way calculations with a square state of size $~8,600 \times 8,600$ with similar results.
There are plenty of misspellings (low-frequency words) in the corpus and the most frequent word is ``the'', which occurs nearly 740,000 times.
At second place is ``be'' with just over 420,000 occurrences.
The task consists of eighty TOEFL synonym tests, which contains five words each.
One example of a synonym test considered here is presented in \tab{toefl}.
\begin{table}[h]
\caption{Example of a TOEFL synonym test.
The first word is given and the task is to determine which of the four remaining words that is a synonym of that word.
Illustrated are also the number of occurrences of each word in the TASA (Touchstone Applied Science Associates, Inc.) corpus.}
\vspace{1ex}
\centering
\begin{tabular}{ c | c }
\hline
Word & Number of occurrences \\
\hline
essential (given) & 855 \\
basic     & 1920 \\
ordinary  & 837 \\
eager     & 480 \\
possible  & 3348 \\
\hline
\end{tabular}
\label{tab:toefl}
\end{table} 
One out of the five words in each test is given and the task is to identify the synonym of that word among the other four words.
There is only one correct synonym in each case, and consequently three incorrect possibilities.
The task to identify the correct synonym is addressed with the NRI-encoded co-occurrence matrix in the following way.
First, the word--word correlation frequencies of the five words in each synonym test are decoded and sorted in descending order (using the function ``find'' in RITensor).
This means that the first (last) word has the highest (lowest) correlation and therefore is most (least) significant.
In practice the non-orthogonality of index vectors cause interference that makes this top-list an approximate one, with decreasing accuracy further down the list.
We therefore select a subset of the high-score word correlations and base the similarity tests on this subset, omitting other word--word correlations with lower decoded weight.
The set of high-score words that correlate with the given word of a synonym test is denoted with $W_0$ and the sets of high-score words that correlate with the four alternative answers are denoted with $W_1$, $W_2$, $W_3$ and $W_4$.
The similarity of two words is measured with the Jaccard index,
\beq
	J(W_0,W_i) = \frac{|W_1\cap W_i|}{|W_1 \cup W_i|},~~~~i\in [1,4].
	\label{eq:jaccard}
\eeq
This means that two words that share many similar word--word correlations are considered to be similar.
The synonym is identified as the pair of words with the highest value of the Jaccard index out of the four calculated values.
We repeated each test ten times with different sets of index vectors and calculated the average success rate and the standard deviation.
The result is presented in \fig{toefl}.

From these results it is clear that high-frequency words cause significant interference that has a negative effect on the decoding accuracy.
In particular the two-way result with unmodified word--word correlation frequencies is near the noise threshold of $25$ percent correct synonyms, which is the average score expected from random guesses.
The situation improves significantly when the dominance of high frequencies is reduced with a square-root transformation of the co-occurrence weights before encoding.
On this particular task one-way RI outperforms the two-way method, which is expected because the dimension reduction ratio is identical and the only effect of two-way RI is to introduce additional interference between the words.
One benefit of two-way RI is that more words and contexts can be defined incrementally with a minimum impact on the storage requirements, see \sect{ricoding} for further information.
This property is expected to become increasingly interesting for higher-order tensors, but we leave that issue to a future study since the computational requirements are demanding.
For comparison we calculate also the success rate using the traditional one-way RI approach where word similarity is determined with the cosine of angle between full-length context vectors \cite{kanerva_2000,sahlgren_2005}.
The best result obtained with the Jaccard index comparison is $54\pm 3.9$ percent correct synonyms and the corresponding cosine result is $51\pm 3.7$ percent.
This is an intuitively expected result, because the full-length context vectors include noise that to some extent is excluded in the limited high-frequency lists, and only the significant word--word correlations are meaningful.
It is not clear how to find the optimal length of the top-list, see \fig{toefl}, and we expect that it varies.
One possibility is to further develop the statistical understanding of the top-list\eg as illustrated in \fig{toplist} and \fig{toplist_fo}.
This requires an analysis of the fidelity of decoded tensor components\eg by generalizing the developments in \cite{SDM_1988}.

%%%%%%%%%%%%%%%%%%%%%%%%%%%%%%%%%%%%%%%%%%%%%%%%%%%%%%%%%%%%%%%%%%%%%%%%%%%%%%

\section{Conclusions}

Random indexing (RI) is a useful incremental dimension reduction technique that has been successfully applied for a decade in vector space models of natural language.
It has recently been applied in a number of other contexts and the method is generalized here to matrices and higher-order tensors, which means that the method can be applied to new classes of complex problems.
One example is natural language processing, where the input data can have both high rank and (very) high dimensionality, and where typically only a fraction of the actual feature space is informative and useful.
There are a number of obvious ways to formulate and utilize high-rank input data for natural language processing applications\eg
by augmenting standard co-occurrence matrices with temporal information (\cf Temporal RI \cite{Jurgens_2009})
or linguistic relations \cite{baroni:lenci,cruys},
or by incorporating structural information in distributed representations \cite{clark:pulman,yeung:tsang}.
However, there have been few attempts at extending traditional matrix-based natural language processing methods to tensors.
One reason for this is the considerable computational costs involved in working with tensors, and this is something that N-way RI (NRI) is likely to facilitate.
The explicit mathematical formulation of NRI that is presented here serves also as a starting point for further theoretical developments\eg concerning storage capacity and decoding accuracy.
In particular, the analytical result for the approximate orthogonality of ternary vectors, see \fig{dotp}, is useful for that purpose and has not been published before as far as we know.
Representations of information with NRI, either in the traditional way as vectors or in the form of higher-order tensors, is most accurate when the encoded features are sparse and of comparable magnitude.
The presence of a noisy or otherwise non-significant background of information that makes the representation non-sparse does not pose a problem as long as the magnitude of these components is significantly lower than that of the interesting features.
An amplitude ratio of one order of magnitude can be sufficient if the density of significant features is low.
We note that a low density of features can result in a first-order transition from significant decoded features to noise, and that this transition becomes continuous when the density of features increases or the difference of magnitudes between features and noise is low.
This property is interesting since it can allow for an accurate discrimination between signal and noise in some problems. 
The possibility to represent non-sparse tensors in a compact representation that can be incrementally and efficiently encoded is one major advantage of NRI over other methods.
Other appealing features are the possibility to extend the size of the tensor at the insignificant cost of generating new sparse random vectors and that decoding is computationally efficient.
The NRI method depends on high dimensionality of the index vectors and is therefore suitable for complex problems only.
We note that there is an interesting scaling relation between the dimension reduction ratio, $\xi$, the rank, $r$, and size, $N$, of the tensor, and the dimensionality of index vectors, $n$: $n = N \xi^{-1/r}$.
This suggests that high-rank tensors may suffer less from high dimension reduction ratios than vectors and matrices, but that is a matter of further investigation.
In this work we simulate two-way and one-way RI\ie representations of matrices and sets of vectors, respectively.
The computationally more demanding task to study the behaviour of higher-order RI is an issue for future work and applications.
Note, however, that the prototype software \cite{RITensor} supports NRI of tensors with any rank.
Our simulation results show that the performance of NRI depends significantly on these parameters: the density of features and their relative magnitude, the dimension reduction ratio and the dimensionality of the index vectors.
The comparison between one-way RI of a set of vectors and two-way RI of matrices shows that the signal to noise ratio of decoded features is higher for one-way RI, when all comparable parameters are equivalent.
This is to be expected, because the two-way algorithm introduces additional interference, practically between the vectors of the one-way method.
The benefit of two-way RI is that the size of the matrix can be extended in both directions by generating additional index vectors.
It would be interesting to know how this trend continues at higher rank of the tensor.
A systematic study of that aspect requires a distributed computing approach and has so far not been attempted by us.
An unexpected result is that the average performance of NRI does not depend significantly on the dimensionality of the index vectors, provided that the dimensionality exceeds a critical threshold of about $n=10^3$.
Below the threshold the average number of correctly decoded features decreases notably with decreasing dimensionality, but above the threshold the average is practically independent of dimensionality.
The standard deviation of the number of correctly decoded features is, however, significantly dependent on the dimensionality.
When increasing the dimensionality of index vectors the standard deviation decreases.
We find that the number of non-zero trits in the index vectors, $\chi_{\cal D}$ ($=2k$), has an effect on the performance but that the effect is smaller than expected from it's impact on the approximate orthogonality of index vectors.
The performance increases notably when increasing $\chi_{\cal D}$ from two to four, but there is apparently no practical reason to go beyond $\chi_{\cal D}=8$.
This indicates that there is a tradeoff between the indifference of index vectors and the magnitude of interference between different tensor components.
Further theoretical work is needed to clarify these findings.

%%%%%%%%%%%%%%%%%%%%%%%%%%%%%%%%%%%%%%%%%%%%%%%%%%%%%%%%%%%%%%%%%%%%%%%%%%%%%%

%\acknowledgments
%\section*{Acknowledgments}

%%%%%%%%%%%%%%%%%%%%%%%%%%%%%%%%%%%%%%%%%%%%%%%%%%%%%%%%%%%%%%%%%%%%%%%%%%%%%%

\bibliographystyle{abbrvurl.bst}
\bibliography{refs}

\end{document}